\documentclass{aa}  
\usepackage{graphicx}
\usepackage{txfonts}
\usepackage{natbib}
\begin{document}
   \title{Chemical abundance analysis of the Open Clusters
   Cr~110, \\ NGC~2099 (M~37), NGC~2420, NGC~7789 and M~67 (NGC 2682)
   \thanks{Based on data collected with the fiber spectrograph FOCES
   	  at the 2.2m Calar Alto Telescope. Also based on data from
   	  the 2MASS survey and the WEBDA, VALD, NIST and GEISA
   	  online databases.}}

   %\subtitle{I. Overviewing the $\kappa$-mechanism}
   
   \titlerunning{Abundances of five Open Clusters}

   \offprints{E. Pancino}

   \author{E. Pancino\inst{1} 
      \and R. Carrera\inst{1,2} 
      \and E. Rossetti\inst{1}
      \and C. Gallart\inst{2}}

   \institute{INAF -- Osservatorio Astronomico di Bologna,
              via Ranzani 1, I-40127 Bologna, Italy\\
             \email{elena.pancino@oabo.inaf.it, emanuel.rossetti@oabo.inaf.it,
	     ricardo.carrera@oabo.inaf.it}
         \and
              Instituto de Astrofisica de Canarias,
	      via Lactea s/n, E-38200, La Laguna, Tenerife, Spain\\
             \email{rcarrera@iac.es,carme@iac.es}
	      }

   \date{Received July 24, 2009; accepted October 2, 2009}

   \abstract{The present number of Galactic Open Clusters that have high
   resolution abundance determinations, not only of [Fe/H], but also of other key
   elements, is largely insufficient to enable a clear modeling of the Galactic
   Disk chemical evolution. 
   }
   {To increase the number of Galactic Open Clusters with high quality
   measurements.  
   }
   {We obtained high resolution (R$\sim$30\,000), high quality (S/N$\sim$50-100
   per pixel), echelle spectra with the fiber spectrograph FOCES, at Calar Alto,
   Spain, for three red clump stars in each of five Open Clusters. We used the
   classical Equivalent Width analysis method to obtain accurate abundances of
   sixteen elements: Al, Ba, Ca, Co, Cr, Fe, La, Mg, Na, Nd, Ni, Sc, Si, Ti, V,
   Y. We also derived the oxygen abundance through spectral synthesis of the
   6300~\AA\  forbidden line.
   }
   {Three of the clusters were never studied previously with high resolution
   spectroscopy: we found [Fe/H]=+0.03$\pm$0.02 ($\pm$0.10)~dex for Cr~110;
   [Fe/H]=+0.01$\pm$0.05 ($\pm$0.10)~dex for NGC~2099 (M~37) and
   [Fe/H]=--0.05$\pm$0.03 ($\pm$0.10)~dex for NGC~2420. This last finding is
   higher than typical recent literature estimates by 0.2--0.3~dex approximately
   and in better agreement with Galactic trends. For the remaining clusters, we
   find: [Fe/H]=+0.05$\pm$0.02 ($\pm$0.10)~dex for M~67 and [Fe/H]=+0.04$\pm$0.07
   ($\pm$0.10)~dex for NGC~7789 . Accurate (to $\sim$0.5~km~s$^{-1}$) radial
   velocities were measured for all targets, and we provide the first high
   resolution based velocity estimate for Cr~110,
   $<$$V_r$$>$=41.0$\pm$3.8~km~s$^{-1}$.
   }
   {With our analysis of the new clusters Cr~110, NGC~2099 and NGC~2420, we
   increase the sample of clusters with high resolution based abundances by 5\%.
   All our programme stars show abundance patterns which are typical of open
   clusters, very close to solar with few exceptions. This is true for all the
   iron-peak and s-process elements considered, and no significant
   $\alpha$-enhancement is found. Also, no significant sign of
   (anti-)correlations for Na, Al, Mg and O abundances is found. If
   anticorrelations are present, the involved spreads must be $<$0.2~dex. We
   then compile high resolution data of 57 OC from the literature and we find a
   gradient of [Fe/H] with Galactocentric Radius of
   --0.06$\pm$0.02~dex~kpc$^{-1}$, in agreement with past work and with Cepheids
   and B stars in the same range. A change of slope is seen outside
   $R_{\rm{GC}}$=12~kpc and [$\alpha$/Fe] shows a tendency of increasing with
   $R_{\rm{GC}}$. We also confirm the absence of a significant Age-Metallicity
   relation, finding slopes of --2.6$\pm$1.1~10$^{-11}$~dex~Gyr$^{-1}$ and
   1.1$\pm$5.0~10$^{-11}$~dex~Gyr$^{-1}$ for [Fe/H] and [$\alpha$/Fe]
   respectively. 
   }
   

   \maketitle
%
%__________________________________________________________________

\section{Introduction}

Open clusters (hereafter OC) are the ideal {\em test particles} in the study of
the Galactic disk, providing chemical and kinematical information in different
locations and at different times. Compared to field stars, they have the obvious
advantage of being coeval groups of stars, at the same distance and with a
homogeneous composition. Therefore, their properties can be determined with
smaller uncertainties. Several attempts have been done in the past to derive two
fundamental relations using OC: the {\em metallicity gradient} along the disk
and the {\em age-metallicity relation} (hereafter AMR) of the disk
\citep[e.g.][]{jan79,pan80,tw97,friel02,che03,sal04}, but they were hampered by
the lack of large and homogeneous high quality datasets.

%%%%%%%%%%%%%%%%%%%%%%%%%%%%%%%%%%%%%%%%%%%%%%%%%%%% Observing Logs
\begin{table*}
\begin{minipage}[t]{17.5cm}
\caption{Observing Logs and Programme Stars Information.}             
\label{obslog}      
\centering          
\renewcommand{\footnoterule}{}  
\begin{tabular}{l l c c c c c c c c c c c }     
\hline\hline       
Cluster      & Star & $\alpha_{J2000}$ & $\delta_{J2000}$ &  B   &  V   &I$_{C}$&  R   & K$_S$& $n_{exp}$ & $t_{exp}^{(tot)}$ & S/N($\simeq$6000\AA) \\
             &      & (hrs)            & (deg)            &(mag) &(mag) & (mag) &(mag) &(mag) &           & (sec)             &               \\
\hline                      
Cr~110\footnote{Star names and I$_C$ \& R magnitudes from \citet{daw98}; 
                coordinates and B \& V magnitudes from \citet{bra03}; 
		K$_S$ magnitudes from 2MASS}
             & 2108 & 06:38:52.5       & +02:01:58.4	  &14.79 &13.35 &  ---  & ---  & 9.76 & 6	  & 16200             & 70     \\  
             & 2129 & 06:38:41.1       & +02:01:05.5	  &15.00 &13.66 & 12.17 &12.94 &10.29 & 7 	  & 18900             & 70     \\  
             & 3144 & 06:38:30.3       & +02:03:03.0	  &14.80 &13.49 & 12.04 &12.72 &10.19 & 6	  & 16195             & 65     \\  
NGC~2099 (M~37)\footnote{Star names from \citet{vanz21}; 
                  coordinates from \citet{kiss01}; 
		  B \& V magnitudes from \citet{kalirai01}; 
		  I$_C$ magnitudes from \citet{nila02}; 
		  K$_S$ magnitudes from 2MASS} 
             & 067  & 05:52:16.6       & +32:34:45.6	  &12.38 &11.12 &  9.87 & ---  & 8.17 & 3	  &  3600             & 95     \\
             & 148  & 05:52:08.1       & +32:30:33.1	  &12.36 &11.09 &  ---  & ---  & 8.05 & 3	  &  3600             & 105    \\  
             & 508  & 05:52:33.2       & +32:27:43.5	  &12.24 &10.98 &  ---  & ---  & 7.92 & 3	  &  3900             & 85     \\  
NGC~2420\footnote{Star names from \citet{can70}; 
                  coordinates from \citet{stet00} and \citet{las90}; 
		  B \& V magnitudes from \citet{tw90}; 
		  I$_C$ \& R magnitudes from \citet{stet00}; 
		  K$_S$ magnitudes from 2MASS} 
             & 041  & 07:38:06.2       & +21:36:54.7	  &13.75 &12.67 & 11.61 &12.13 &10.13 & 5	  &  9000             & 70     \\
             & 076  & 07:38:15.5       & +21:38:01.8	  &13.65 &12.66 & 11.65 &12.14 &10.31 & 5 	  &  9000             & 75     \\  
             & 174  & 07:38:26.9       & +21:38:24.8	  &13.41 &12.40 &  ---  & ---  & 9.98 & 5	  &  9000             & 60     \\  
NGC~2682 (M~67)\footnote{Star names from \citet{fag06}; 
              coordinates from \citet{fan96}; 
	      B, V \& I$_C$ magnitudes from \citet{san04};
	      B magnitude for star 286 and R magnitudes from \citet{jan84};
	      K$_S$ magnitudes from 2MASS}   
             & 0141 & 08:51:22.8       & +11:48:01.7	  &11.59 &10.48 &  9.40 & 9.92 & 7.92 & 3 	  &  2700             & 85     \\
             & 0223 & 08:51:43.9       & +11:56:42.3	  &11.68 &10.58 &  9.50 &10.02 & 8.00 & 3 	  &  2700             & 85     \\  
             & 0286 & 08:52:18.6       & +11:44:26.3	  &11.53 &10.47 &  9.43 & 9.93 & 7.92 & 3 	  &  2700             & 105    \\  
NGC~7789\footnote{Star names and V \& I$_c$ magnitudes from \citet{gim98b}; 
                  J1950 coordinates from \citet{kun23}, precessed to J2000; 
		  B magnitudes from \citet{moc99}; 
		  K$_S$ magnitudes from 2MASS} 
             & 5237 & 23:56:50.6       & +56:49:20.9	  &13.92 &12.81 & 11.52 & ---  & 9.89 & 5 	  &  9000            &  70   \\
             & 7840 & 23:57:19.3       & +56:40:51.5	  &14.03 &12.82 & 11.49 & ---  & 9.83 & 6	  &  9000            &  75   \\  
             & 8556 & 23:57:27.6       & +56:45:39.2	  &14.18 &12.97 & 11.65 & ---  &10.03 & 3	  &  5400            &  45   \\  
\hline 
\end{tabular}
\end{minipage}
\end{table*}
%%%%%%%%%%%%%%%%%%%%%%%%%%%%%%%%%%%%%%%%%%%%%%%%%%%% Observing Logs

In particular, the lack of a {\em metallicity scale} extending to solar
metallicity with comparable precision to the lower metallicity regimes
\citep[i.e.,][]{zin84,car97} represents the main problem from the point of view of {\em
(i)} the study of the Galactic disk; {\em (ii)} tests of stellar evolution
models for younger and more metal-rich {\em simple stellar populations} and {\em
(iii)} the use of those stellar populations as templates for extragalactic
studies of population synthesis. Of the $\sim$1700 known OC \citep[][and
updates]{dia02}, only a subset of $\sim$140, i.e., 8\% of the total, possesses
some metallicity determination. Most of these have been obtained through
different photometric studies in the Washington \citep[e.g.][]{gei91,gei92}, DDO
\citep[e.g.][]{cla99}, Str\"omgren \citep[e.g.][]{bru99,twa03}, UBV
\citep[e.g.][]{cam85} and IR \citep[e.g.][]{tie97} photometric systems and
passbands, giving often rise to considerable differences with those obtained
from spectroscopy \citep[see][and re\-fe\-ren\-ces therein]{gra00}. In a much
smaller number of clusters, abundances have been derived from low-resolution
spectroscopy \citep[e.g.,][]{ricardo,war08}, with admirable attempts to obtain
large and homogeneous datasets \citep[see][]{friel93,friel02}, in spite of the
non-negligible uncertainties involved in the procedure. 

A few research groups (see Section~\ref{sec-disc} for more details) are
presently obtaining high quality spectra and are producing more precise
abundance determinations. The study of elements other than the iron-peak ones
(such as $\alpha$, s- and r-process, light elements), allows one to put more
constraints on the sites of production of those elements (SNe Ia, SNe II, giants
and supergiants, Wolf-Rayet stars), and therefore on their production
timescales. These are fundamental ingredients for the chemical evolution
modeling of the Galactic Disk \citep{tos82,chi01,col09}.

For these reasons, we obtained high resolution spectra for a sample of poorly
studied old OC. We present here the detailed abundance analysis of five clusters
observed during our first run at Calar Alto. Observations and data reductions
are described in Section~\ref{sec-obs}; the linelist and equivalent width
measurements are detailed in Section~\ref{sec-li} while the abundance analysis
methods and results are presented in Section~\ref{sec-abo}; abundance results
are then discussed and compared with literature results in
Sections~\ref{sec-lit}, \ref{sec-disc} and \ref{sec-trend}; finally, we summarize
our results and draw our conclusions in Section~\ref{sec-sum}.

%__________________________________________________________________

\section{Observational Material}
\label{sec-obs}

\begin{figure*}
\centering
\includegraphics[bb=40 40 590 280,clip]{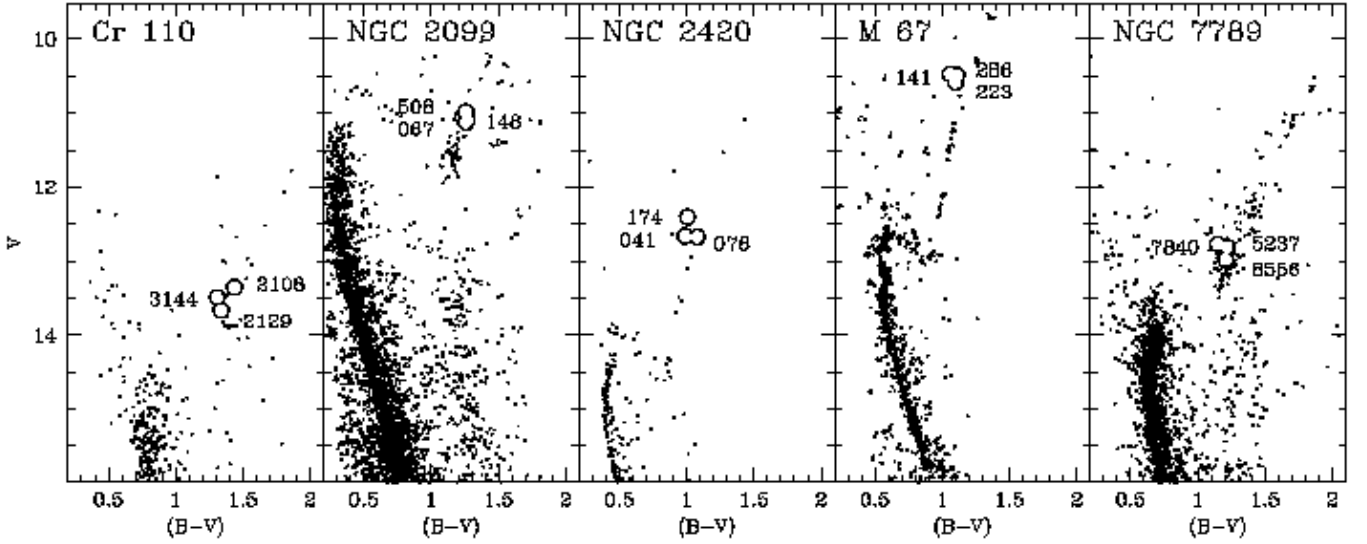}

\caption{V, (B--V) Color Magnitude Diagrams of the programme clusters
(from the WEBDA), with the location of our target stars.}

\label{cmds}
\end{figure*}

Three red clump stars\footnote{Though fainter than the brigthest giants, red
clump stars have the advantage of a higher gravity and temperature, that reduces
considerably line crowding. Also, clump stars are easy to identify even in the
sparsest cluster, maximizing the chance of choosing cluster members.} were
selected in each of the target clusters using the WEBDA\footnote{{\tt
http://www.univie.ac.at/webda}} database \citep{webda} and the
2MASS\footnote{{\tt http://www.ipac.caltech.edu/2mass}. 2MASS (Two Mi\-cron All
Sky Survey) is a joint project of the University of Massachusetts and the
Infrared Processing and Analysis Center/California Institute of Technology,
funded by the National Aeronautics and Space Administration and the National
Science Foundation.} survey data for the infrared K$_S$ magnitudes
\citep{2mass2}. More details on references for star names, coordinates and
magnitudes can be found in Table~\ref{obslog}, while the position of our targets
in the Color Magnitude Diagrams (CMDs) obtained from WEBDA are shown in
Figure~\ref{cmds}.

%%%%%%%%%%%%%%%%%%%%%%%%%%%%%% %%%%%%%%%%%%%%%%%%%%%% DAOSPEC Errors
%%%%%%%%%%%%%%%%%%%%%%%%%%%%%%%%%%%%%%%%%%%%%%%%%%%% DAOSPEC Errors

Observations were done between the 1st and 10th of January 2004 with the fiber
echelle spectrograph FOCES at the 2.2~m Calar Alto Telescope, in Spain. The sky
was generally clear, though a few nights had thin cirrus and sometimes clouds,
forcing us to increase the exposure times considerably. All stars were observed
in 3--16 exposures lasting 15--90~min each, depending on the magnitude, until a
global S/N ratio between 70 and 100 (per pixel) was reached around 6000~\AA\ 
(Table~\ref{obslog}). Each night we took a sky exposure lasting as the longest
one of the night. The sky level was negligible for all exposures with S/N$>$20,
therefore exposures with S/N$<$20 were neglected and we did not subtract the
sky, to avoid adding noise to the spectra. Sky emission lines, even in the red
part of the spectrum, did only occasionally disturb the measurements of some
absorption lines, that were discarded. The spectral resolution was $R\simeq30\,
000$ for all spectra. 

\subsection{Data Reductions}

Data reductions were done with IRAF\footnote{Image Reduction and Analysis
Facility. IRAF is distributed by the National Optical Astronomy Observatories,
which is operated by the association of Universities for Research in Astronomy,
Inc., under contract with the National Science Foundation} within the {\em imred}
and {\em echelle} packages. The following steps were performed: bias subtraction,
flatfielding, order tracing with optimal extraction, background subtraction,
wavelength calibration with the help of a Thorium-Argon lamp, and final merging
(and rebinning) of overlapping orders. The one-dimensional spectra obtained from
different exposures (with S/N$>$20) were median-averaged to produce one single
high S/N spectrum for each star, used for equivalent width measurements
(Section~\ref{sec-ew}). Finally, the noisy ends of each combined spectrum were
cut, allowing for an effective wavelength coverage from 5000 to 9000~\AA.

Sky absorption lines (telluric bands of O$_2$ and H$_2$O) were removed using the
IRAF task {\em telluric} with the help of two hot, rapidly rotating stars,
HR~3982 and HR~8762, chosen from the {\em Bright Star Catalogue} \citep{bsc}.
HR~3982 and HR~8762 were observed each night at an airmass not too different from
the scientific targets. Residuals of the correction in the red part of the
spectrum (for example from the strong O$_2$ band around 7600~\AA) prevented us
from using most of the corresponding spectral regions for our abundance analysis.
Also, after 8400~\AA, the echelle orders do not overlap anymore and small gaps
appear.

%__________________________________________________________________

\subsection{Radial Velocities}
\label{sec-vel}

Radial velocities were measured with the help of DAOSPEC \citep[][see also
Section~\ref{sec-ew}]{daospec}. Measurements were based on $\simeq$360
absorption lines of different elements (see Section~\ref{sec-li}) with typical
measurement errors on the mean of about 0.1~km~s$^{-1}$. All measurements were
performed separately on the one-dimensional spectra extracted from the single
exposures for each star, including those with S/N$<$20, that were not used for
the abundance analysis. In this way, we could check that no significant radial
velocity variations were present.

%%%%%%%%%%%%%%%%%%%%%%%%%%%%%%%%%%%%%%%%%%%%%%%%% Radial Velocities
\begin{table}
\begin{minipage}[htb]{\columnwidth}
\caption{Heliocentric radial velocities measurements and 1$\sigma$ errors
($V_r \pm \delta V_r$)$_{here}$ for each programme star. Literature
measurements, when available, are also reported with their uncertainties
($V_r \pm \delta V_r$)$_{lit}$.}             
\label{radvel}      
\centering          
\renewcommand{\footnoterule}{}  
\begin{tabular}{l l c c c c}     
\hline\hline       
%HJD & $E$ & Method\#2 & \multicolumn{4}{c}{Method\#3}\\ 
Cluster  & Star & $(V_r\pm\delta V_r)_{here}$ & $(V_r\pm\sigma V_r)_{lit}$ \\
         &      &(km~s$^{-1}$)       &(km~s$^{-1}$) \\
\hline           
Cr~110   & 2108 &   44.47 $\pm$ 0.54 & $45\pm8$\footnote{Cluster average 
         by \citet{ricardo}, based on 8 stars.} \\  
         & 2129 &   38.74 $\pm$ 0.64 & $45\pm8$$^a$ \\  
         & 3144 &   37.81 $\pm$ 0.69 & $45\pm8$$^a$ \\  
NGC~2099 (M~37)
         & 067  &    8.79 $\pm$ 0.10 &   8.04 $\pm$ 0.19\footnote{\citet{mer08},
	 superseding \citet{mer96}.}    \\
         & 148  &    9.05 $\pm$ 0.36 &   8.73 $\pm$ 0.19$^b$ \\  
         & 508  &    9.40 $\pm$ 0.28 &   8.96 $\pm$ 0.19$^b$ \\  
NGC~2420
         & 041  &   74.23 $\pm$ 0.87 &   79.0 $\pm$ 4.2\footnote{Average of 
	 measurements by \citet{liu87}, \citet{friel89} 
         and \citet{scott95}.}  \\
         & 076  &   75.49 $\pm$ 0.41 &   75.7 $\pm$ 8.6$^c$ \\  
         & 174  &   73.66 $\pm$ 1.17 &   68.0 $\pm$ 0.9$^c$ \\  
NGC~2682 (M~67)     
	 &  141 &   35.23 $\pm$ 0.05 &   33.5 $\pm$ 2.3\footnote{Average
	 of measurements by \citet{mat86}, \citet{pil88}, 
	 \citet{friel89}, \citet{friel93}, \citet{sun92},
	 \citet{scott95},  \citet{melo01}, \citet{yong05},
	 \citet{yad08}.} \\ 
	 &  223 &   34.92 $\pm$ 0.31 &   33.0 $\pm$ 1.1$^d$ \\  
         &  286 &   38.90 $\pm$ 0.46 &   33.2 $\pm$ 2.0$^d$ \\  
NGC~7789 
         & 5237 & --51.06 $\pm$ 0.99 & --57.17 $\pm$ 0.23\footnote{\citet{gim98a}.}  \\
         & 7840 & --49.06 $\pm$ 0.81 & --49.21 $\pm$ 0.28$^e$ \\  
         & 8556 & --53.37 $\pm$ 0.81 & --54.10 $\pm$ 0.32$^e$ \\  
\hline             	  
\end{tabular}	   
\end{minipage}	   
\end{table}	   
%%%%%%%%%%%%%%%%%%%%%%%%%%%%%%%%%%%%%%%%%%%%%%%%% Radial Velocities

%%%%%%%%%%%%%%%%%%%%%%%%%%%%%%%%%%%%%%%%%%%%%%%% Stellar Parameters
\begin{table*}
\caption{Stellar Parameters for the programme stars.}             
\label{pars}      
\centering          
\begin{tabular}{l l  c c c c c c c c}     
\hline\hline       
%HJD & $E$ & Method\#2 & \multicolumn{4}{c}{Method\#3}\\ 
Cluster         & Star &T$_{\rm{eff}}^{(phot)}$&T$_{\rm{eff}}^{(spec)}$& $\log g^{(phot)}$&$\log g^{(spec)}$&$v_t^{(phot)}$  &$v_t^{(spec)}$  & M$_{clump}$ \\
                &      & (K)                   & (K)	               & (cgs)            & (cgs)           &(km~s$^{-1}$)   &(km~s$^{-1}$)   & (M$_\odot$) \\
\hline     	     
Cr~110          & 2108 & 4914$\pm$111  & 4650 & 2.32$\pm$0.16 & 2.8 & 1.20$\pm$0.02/1.62$\pm$0.04 & 1.4 & 1.9$\pm$0.1 \\  
                & 2129 & 5056$\pm$223  & 4950 & 2.53$\pm$0.19 & 2.7 & 1.17$\pm$0.02/1.52$\pm$0.06 & 1.4 & 1.9$\pm$0.1 \\  
                & 3144 & 5112$\pm$258  & 4800 & 2.49$\pm$0.20 & 2.8 & 1.18$\pm$0.02/1.49$\pm$0.08 & 1.3 & 1.9$\pm$0.1 \\  
NGC~2099 (M~37) & 067  & 4773$\pm$119  & 4550 & 2.15$\pm$0.32 & 2.7 & 1.22$\pm$0.03/1.68$\pm$0.05 & 1.5 & 2.7$\pm$0.2 \\
                & 148  & 4708$\pm$116  & 4550 & 2.10$\pm$0.31 & 2.7 & 1.28$\pm$0.03/1.72$\pm$0.04 & 1.5 & 2.7$\pm$0.2 \\  
                & 508  & 4715$\pm$115  & 4500 & 2.05$\pm$0.32 & 2.8 & 1.23$\pm$0.03/1.72$\pm$0.04 & 1.5 & 2.7$\pm$0.2 \\  
NGC~2420        & 041  & 4616$\pm$77   & 4850 & 2.43$\pm$0.06 & 2.6 & 1.19$\pm$0.01/1.76$\pm$0.03 & 1.4 & 1.4$\pm$0.1 \\
                & 076  & 4755$\pm$103  & 4800 & 2.51$\pm$0.06 & 2.6 & 1.18$\pm$0.01/1.69$\pm$0.03 & 1.6 & 1.4$\pm$0.1 \\  
                & 174  & 4730$\pm$77   & 4800 & 2.39$\pm$0.05 & 2.6 & 1.19$\pm$0.01/1.71$\pm$0.03 & 1.5 & 1.4$\pm$0.1 \\  
NGC~2682 (M~67) & 141  & 4590$\pm$100  & 4650 & 2.42$\pm$0.04 & 2.8 & 1.19$\pm$0.01/1.78$\pm$0.04 & 1.3 & 1.3$\pm$0.1 \\
                & 223  & 4594$\pm$100  & 4800 & 2.46$\pm$0.04 & 2.8 & 1.18$\pm$0.00/1.78$\pm$0.04 & 1.3 & 1.3$\pm$0.1 \\  
                & 286  & 4653$\pm$103  & 4850 & 2.45$\pm$0.04 & 2.8 & 1.18$\pm$0.01/1.75$\pm$0.04 & 1.4 & 1.3$\pm$0.1 \\  
NGC~7789        & 5237 & 4868$\pm$168  & 4900 & 2.53$\pm$0.15 & 2.8 & 1.17$\pm$0.02/1.63$\pm$0.04 & 1.2 & 1.8$\pm$0.1 \\
                & 7840 & 4759$\pm$131  & 4800 & 2.47$\pm$0.15 & 2.7 & 1.18$\pm$0.02/1.69$\pm$0.04 & 1.5 & 1.8$\pm$0.1 \\  
                & 8556 & 4775$\pm$135  & 4900 & 2.54$\pm$0.15 & 2.9 & 1.17$\pm$0.02/1.68$\pm$0.04 & 1.4 & 1.8$\pm$0.1 \\  
\hline                  
\end{tabular}
\end{table*}
%%%%%%%%%%%%%%%%%%%%%%%%%%%%%%%%%%%%%%%%%%%%%%%% Stellar Parameters

Heliocentric corrections were computed with the IRAF task {\em rvcor},
which bears a negligible uncertainty of less than 0.005~km~s$^{-1}$. Since
we did not observe any radial velocity standard and our calibration lamps
were not taken simultaneously, we used telluric absorption lines to find
the absolute zero point of our radial velocity measurements. In
particular, laboratory wavelengths of the H$_2$O absorption bands around
5800, 6500, 7000, 7200, 8000 and 8900~\AA\  and of the O$_2$ absorption
bands around 6300, 6900 and 7600~\AA\  were obtained from the
GEISA\footnote{{\tt http://ara.lmd.polytechnique.fr/htdOC-public/pro
ducts/GEISA/HTML-GEISA/}} database \citep{geisa1,geisa2} and we measured
their radial velocity in our programme stars. The resulting zeropoint
corrections, based on 200--250 telluric lines, amount generally to no more
than $\pm$1~km~s$^{-1}$ with typical errors on the mean of about
0.5~km~s$^{-1}$, approximately five times larger than those on the radial
velocity measurements.

After applying the above corrections, and propagating the corresponding
uncertainties, we computed a weighted average of the heliocentric velocities
estimates for each exposure (see Table~\ref{radvel}). All the programme stars
appear to be radial velocity members of the observed clusters, with the possible
exception of star 2108 in Cr~110, that has a slightly higher velocity than 2129
and 3144. However, since the value for 2108 is within 3$\sigma$ from the mean
value for the cluster, we decided not to reject this star. We can provide the
first radial velocity estimate based on high resolution for Cr~110:
$<$$V_r$$>$=41.0$\pm$3.8~km~s$^{-1}$. Our determinations are generally in good
agreement with literature values within 3$\sigma$, except maybe for star 5237 in
NGC~7789, which is marginally dicrepant with the estimate by \citet{gim98a}.
However, there is perfect agreement between the two studies for the other two
stars of NGC~7789, and our estimate appears more in line with a membership of
5237. In conclusion, we considered all the programme stars as likely radial
velocity members of their respective clusters.

%__________________________________________________________________

\subsection{Photometric Parameters}
\label{sec-phot}

We first computed the dereddened colors\footnote{Since R magnitudes are available
for less than half of our sample, we decided to ignore them in the following.}
(B--V)$_0$, (V--I$_C$)$_0$\footnote{After dereddening, (V--I$_C$) was also
converted into (V--I$_J$) using the prescription by \citet{bes79}, to be used with
the color-temperature relations by \citet{alo99}.} and
(V--K$_TCS$)$_0$\footnote{We computed the K$_TCS$ magnitudes from the 2MASS TCS
magnitudes using the prescription by \citet{kin02}.}. The adopted E(B--V) values
are indicated in Table~\ref{phot}, where E(V--I$_C$) was obtained with the
reddening laws by \citet{dean78}, and E(V--K$_TCS$) with \citet{car89}. We were
then able to obtain T$_{\rm{eff}}$ and the bolometric correction BC$_{\rm{V}}$ for
each programme star, using both the \citet{alo99} and the (theoretical and
empyrical) \citet{mon98} color-temperature relations, taking into account the
uncertainties on magnitudes and reddening estimates. The average difference
between the \citet{alo99} and the \citet{mon98} temperatures was $\Delta
T_{\rm{eff}}$$=$178$\pm$66~K \citep[for the empyrical calibration of][]{mon98} and
$\Delta T_{\rm{eff}}$$=$127$\pm$69~K \citep[for the theoretical calibration
of][]{mon98}. We averaged all the above $T_{\rm{eff}}$ estimates to obtain our
photometric reference values with their 1$\sigma$ uncertainties
(Table~\ref{pars}).

Gravities were obtained from T$_{\rm{eff}}$ and BC$_V$ using the fundamental
relations

\begin{eqnarray}
\nonumber
\log \frac{g}{g_\odot}=\log \frac{M}{M_\odot}+2\log
\frac{R_\odot}{R}\\
\nonumber
0.4 (M_{\rm{bol}}-M_{\rm{bol},\odot})=-4\log
\frac{T_{\rm{eff}}}{T_{\rm{eff},\odot}}+2\log \frac{R_\odot}{R}
\end{eqnarray}

\noindent where red clump masses were derived using Table~1 of \citet{gir01},
and are also shown in Table~\ref{pars}. We assumed $\log g_{\odot}=4.437$,
$T_{\rm{eff},\odot}=5770~\rm{K}$ and $M_{\rm{bol},\odot}=4.75$, in conformity
with the IAU recommendations \citep{iau}. The difference between the
\citet{alo99} and the (empyrical and theoretical, respectively) \citet{mon98}
estimates was $\Delta \log g$$=$0.20$\pm$0.13 and $\Delta \log
g$$=$0.18$\pm$0.13. As above, we averaged all our estimats to obtain photometric
gravities $\log g^{(phot)}$ (Table~\ref{pars}) and their 1$\sigma$
uncertainties.

%%%%%%%%%%%%%%%%%%%%%%%%%%%%%%%%%%%%%%%%%%%%%%%% Cluster Parameters
\begin{table}
\begin{minipage}[htb]{\columnwidth}
\caption{Input cluster parameters.}             
\label{phot}      
\centering          
\renewcommand{\footnoterule}{}  
\begin{tabular}{l c c c }     
\hline\hline       
%HJD & $E$ & Method\#2 & \multicolumn{4}{c}{Method\#3}\\ 
Cluster   & E(B--V) & (m-M)$_V$ & log Age \\
          & (mag)   & (mag)     & (dex) \\
\hline     	     
Cr~110\footnote{Averages of measurements by \citet{tsa71},
      \citet{daw98} and \citet{bra06}.}
          & 0.54$\pm$0.03 & 13.37$\pm$0.38 & 9.23$\pm$0.15 \\	  
NGC~2099 (M~37)\footnote{Averages of measurements by \citet{west67a}, 
      \citet{jen75}, \citet{lynga87}, \citet{mer96}, \citet{tw97}, 
      \citet{kiss01}, \citet{kalirai01}, \citet{nila02}, \citet{gro03}, 
      \citet{bra06} and \citet{har08}.}
          & 0.27$\pm$0.04 & 11.53$\pm$0.19 & 8.61$\pm$0.16 \\
NGC~2420\footnote{Averages of measurements by \citet{mcc74},
      \citet{bar84}, \citet{van85}, \citet{lynga87}, \citet{tw90},
      \citet{car94}, \citet{dem94}, \citet{tw97}, \citet{tad02} and
      \citet{gro03}.} 
          & 0.04$\pm$0.03 & 11.88$\pm$0.27 & 9.47$\pm$0.17 \\
NGC~2682 (M~67)\footnote{Averages of measurements by \citet{rac71},
      \citet{bar84},  \citet{van85}, \citet{nis87}, \citet{hob91},
      \citet{gil92},  \citet{dem92}, \citet{mey93}, \citet{mon93},
      \citet{car94}, \citet{din95}, \citet{fan96}, \citet{tw97},
      \citet{sar99}, \citet{tad02}, \citet{gro03}, \citet{van03},
      \citet{lau04} and \citet{san04}.}     
          & 0.04$\pm$0.02 &  9.67$\pm$0.11 & 9.64$\pm$0.05 \\
NGC~7789\footnote{Averages of measurements by \citet{bur58},
      \citet{arp62}, \citet{str70}, \citet{jen75}, \citet{jan77},
      \citet{cla79}, \citet{twa85}, \citet{lynga87}, \citet{maz88},
      \citet{friel93}, \citet{mar94}, \citet{jan94}, \citet{tw97},
      \citet{gim98a}, \citet{sar99}, \citet{val00}, \citet{tad02}
      and \citet{bar04}.}  
          & 0.27$\pm$0.04 & 12.23$\pm$0.20 & 9.21$\pm$0.12 \\
\hline                  
\end{tabular}
\end{minipage}
\end{table}
%%%%%%%%%%%%%%%%%%%%%%%%%%%%%%%%%%%%%%%%%%%%%%%% Cluster Parameters

Finally, a photometric estimate of the microturbulent velocities $v_t$ was
obtained using both the prescriptions of \citet{ram03},
$v_t=4.08-5.01~10^{-4}~T_{\rm{eff}}$, and of \citet{car04}, $v_t=1.5-0.13~\log
g$. The latter takes into account the systematic effect discussed by
\citet{mag84} and is on average lower by $\Delta v_t=0.49 \pm 0.08$~km~s~$^{-1}$
than the one by \citet{ram03}. However, the actual amount of the correction for
the \citet{mag84} effect depends heavily on the data quality (i.e., resolution,
S/N ratio, number of lines used, log~$gf$ values etc.). Therefore we chose not
to average the two estimates, but to use them as an indication of the (wide)
$v_t$ range to explore in our abundance analysis (see Section~\ref{sec-best}).

%__________________________________________________________________

\section{Linelist, Atomic Data and Equivalent Widths}
\label{sec-li}

\begin{figure}
\centering
\includegraphics[width=\columnwidth]{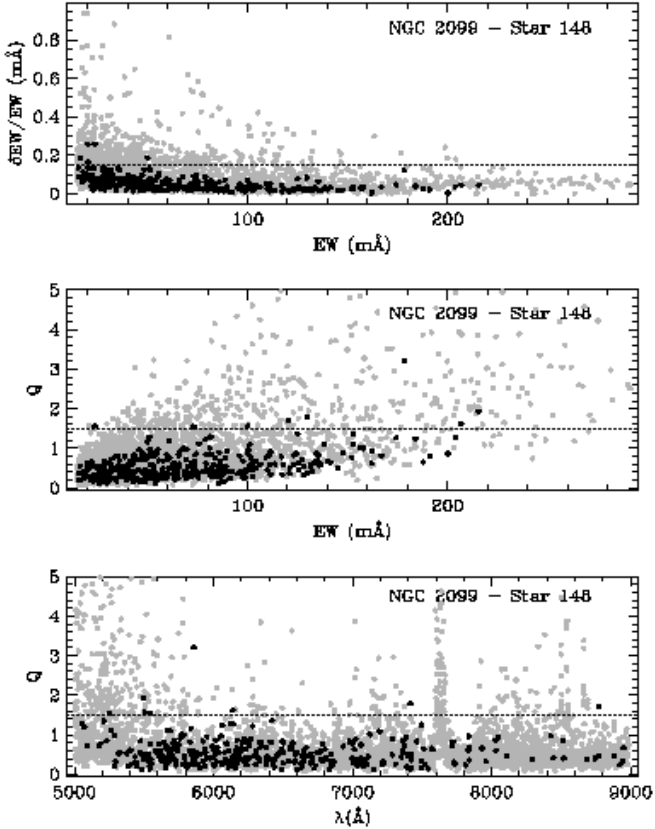}

\caption{The behaviour of DAOSPEC relative errors $\delta$EW$/$EW versus EW is
shown (top panel) and the 15$\%$ error limit is marked with a dotted line. The
behaviour of the quality parameter Q is shown versus EW (middle panel) and
wavelength (bottom panel) and the Q$=$1.5 limit is marked with a dotted line. See
Section~\ref{sec-ew} for more details. In all panels, grey dots represent all the
lines measured by DAOSPEC, while black dots represent lines cross-identified with
our input linelist.}

\label{errors}
\end{figure}

We created a masterlist of absorption lines by visually comparing our spectra
with the the UVES solar spectrum\footnote{{\tt
http://www.eso.org/observing/dfo/quality/UVES/pipel ine/solar\_spectrum.html}}
in the range 5000--9000~\AA, and with the line lists extracted from the
VALD\footnote{{\tt http://www.astro.uu.se/$\sim$vald/}} database \citep{vald}
and the Moore\footnote{{\tt ftp://ftp.noao.edu/fts/linelist/Moore}}
\citep{moore} solar atlas. The masterlist was fed to DAOSPEC and EW were
measured for all our programme stars. A first selection was applied to reject
all those lines that were measured in 10 stars or less (out of 15) and that had
EW systematically larger than 250~m\AA. Later, after performing a first crude
abundance analysis (see Section~\ref{sec-best}), we rejected all the lines that
gave sistematically discrepant abundances, especially if the formal DAOSPEC
relative error ($\delta$EW/EW, Figure~\ref{errors}) was around 15\% or more, and
the DAOSPEC quality parameter was above 1.5 (for more details about DAOSPEC
error and quality parameter, see Section~\ref{sec-ew}). The final linelist,
including atomic data and EW measurements for all programme stars, contains 358
absorption lines of 17 species, and can be found in the electronic version of
Table~\ref{tab-ew}. Atomic data include laboratory wavelengths, excitation
potentials and $\log gf$ values, which are always taken from the VALD database,
with the exceptions listed below.

\subsection{$\alpha$ Elements Atomic Data}
\label{ato-alpha}

The only $\alpha$-element for which we had clear problems with the atomic data
was magnesium. The lines with $\chi_{ex}$=5.75~eV (7060 and 7193\AA) gave
discrepant abundances by $\sim$1.5 dex with respect to the average of all Mg
lines. We compared our VALD log~$gf$ with the NIST\footnote{\tt
http://physics.nist.gov/PhysRefData/ASD/index.html} database of atomic data and
noticed a difference of 1.4 dex for the $\chi_{ex}$=5.75~eV lines, while all the
other Mg~I had very similar log~$gf$ values in both databases. The NIST log~$gf$
values abundances of the $\chi_{ex}$=5.75~eV lines gave abundances in much
better agreement with the other Mg~I lines and with the literature Mg abundances
for OC, therefore we used the NIST values for those lines, instead of the VALD
ones. 

Another element with uncertain log~$gf$ values is Calcium. As an example, for
the 9 lines we use, there is an average difference of
log$gf$$_{NIST}$--log$gf$$_{VALD}$=--0.17$\pm$0.18~dex, which is not
statistically significant given the large $\sigma$. Also, the NIST log$gf$
values for those 9 lines all range from D to E, which means that they are
largely uncertain. Finally, our solar abundance (Section~\ref{sec-sun}) gives
[Ca/H]=--0.09$\pm$0.03 ($\pm$0.03)~dex if we use the VALD log$gf$ and
[Ca/H]=+0.08$\pm$0.03 ($\pm$0.03)~dex with the NIST ones, which is equally
compatible with zero within 3 sigma. Summarizing, there is large uncertainty on
the Calcium log$gf$ determinations, and we really should keep in mind that there
is an additional $\sim$0.2~dex uncertainty on all [Ca/Fe] determinations in the
literature.

For the synthesis of the [O~I]--Ni~I blend at 6300\AA, we used the VALD log~$gf$
for oxygen, but we chose to use the \citet{nickel} log~$gf$ for Ni~I at
6300.35\AA, which is lower (--2.11~dex instead of --1.74) and gives oxygen
abundances more in line with the other $\alpha$-elements.

%%%%%%%%%%%%%%%%%%%%%%%%%%%%%%%%%%%%%%%%%%%%%%%%%%%% DAOSPEC Errors
%%%%%%%%%%%%%%%%%%%%%%%%%%%%%%%%%%%%%%%%%%%%%%%%%%%% DAOSPEC Errors

%%%%%%%%%%%%%%%%%%%%%%%%%%%%%%%%%%%%%%%%%%%%%%%%% Equivalent Widths
\begin{table*}
\caption{Equivalent Widths and Atomic Data of the programme stars. The
complete version of the Table is available at the CDS. Here we show a few lines
for guidance about its contents.}             
\label{tab-ew}      
\centering          
\begin{tabular}{c c c c c c c c c c}     
\hline\hline       
Cluster & Star & $\lambda$ & Elem  & $\chi_{\rm{ex}}$&$\log gf$& EW     &$\delta$EW & Q &  \\
&& (\AA)       &       & (eV)            & (dex)   & (m\AA) & (m\AA)    &   &  \\
\hline     	     
Cr~110 & 2108 & 5055.99 & Fe I & 4.31 & -2.01 & 41.2 & 3.3 & 0.371 \\  
Cr~110 & 2108 & 5178.80 & Fe I & 4.39 & -1.84 & 45.4 & 7.1 & 0.851 \\  
Cr~110 & 2108 & 5294.55 & Fe I & 3.64 & -2.86 &  --  & --  & --    \\
Cr~110 & 2108 & 5285.13 & Fe I & 4.43 & -1.64 & 50.1 & 5.2 & 0.607 \\
Cr~110 & 2108 & 5295.31 & Fe I & 4.42 & -1.69 & 38.3 & 9.5 & 1.958 \\  
\hline \hline                 
\end{tabular}
\end{table*}
%%%%%%%%%%%%%%%%%%%%%%%%%%%%%%%%%%%%%%%%%%%%%%%%% Equivalent Widths

\subsection{Heavy Elements Atomic Data}
\label{ato-heavy}

For neodymium, we could only find three reliable lines, which apparently do not
need any detailed HFS (hyper-fine splitting) analysis \citep{aoki01}, at 5092,
5249 and 5485~\AA. However, the spread of their abundances was quite high
(Table~\ref{abotab}). The laboratory log~$gf$ published by \citet{den03} are
very similar to the ones from VALD, except for the 5485~\AA\  line, where they
differ by 0.14 dex. Therefore, since the log~$gf$ values by \citet{den03}
slightly reduced the spread in the [Nd/Fe], we used them instead of the VALD
ones (see Table~\ref{tab-ew}).

\subsection{Equivalent Widths with DAOSPEC}
\label{sec-ew}

The full description of how DAOSPEC works, including comparisons with
the literature and several experiments with artificial and real spectra,
can be found in \citet{daospec}. The instructions on how to install,
configure and use DAOSPEC can be found in {\em ``Cooking with
DAOSPEC''}\footnote{\tt
http://www3.cadc-ccda.hia-iha.nrc-cnrc.gc.ca/commun ity/STETSON/daospec/
;  http://www.bo.astro.it/$\sim$pancino /projects/daospec.html}. In
short, DAOSPEC is a Fortran program that automatically finds absorption
lines in a stellar spectrum, fits the continuum, measures EW, and
identifies lines with a laboratory linelist; it also provides a radial
velocity estimate (Section~\ref{sec-vel}).  

As described in Section~\ref{sec-li}, we used the DAOSPEC errors and quality
parameter, Q, to select good absorption lines in our master line list. Since our
spectra are rebinne linear in wavelength, we scaled the FWHM with $\lambda$.
Figure~\ref{errors} shows their behaviour. $\delta EW$ is the formal error of
the gaussian fit that DAOSPEC outputs, and $\delta EW/EW$ can be used to select
good measurements, since smaller lines are noisier and tend to have higher
relative errors. The quality parameter Q, instead, is the result of the
comparison of local residuals around each line with average residuals on the
whole spectrum. As a result Q tends to be worse for strong lines, because the
Gaussian approximation does not hold so well anymore. Also, Q gets worse at the
blue side of the spectrum, where the S/N ratio is lower. In the region around
7700~\AA, where the residuals of the prominent O$_2$ telluric band disturb the
measurements, Q reaches its maximum. The measured EW for our programme stars are
shown in the electronic version of Table~\ref{tab-ew} along with the $\delta$EW
and Q parameter estimated by DAOSPEC.

\subsection{EW uncertainties}
\label{sec-cont}

We used the formal errors of the Gaussian fit computed by DAOSPEC only to
reject bad measurements from our initial line list. The actual abundance
errors due to the EW measurements process itself were instead computed later,
as explained in  Section~\ref{sec-err1}. 

To compute the EW uncertainty due to the continuum placement, we used Equation
7 from \citet{daospec} to derive the effective uncertainty on the continuum
placement ($\pm \Delta C / C $), which turned out to be significantly smaller
than 1\%. We first lowered the "best" continuum by $\Delta C/ C$ and measured
EWs again, obtaining EW$_{(-)}$, then we increased it by the same amount and
measured EW$_{(+)}$. The differences with the ``correct'' EW measurements,
$\Delta$EW$_{(-)}$ and $\Delta$EW$_{(+)}$ were averaged to produce the actual
$\Delta EW$ for each line. The typical resulting uncertainty, due only to the
continuum placement, was approximately constant with EW and of
$\Delta$EW$\simeq$1~m\AA\  approximately \citep[see also Figure~2
by][]{daospec}. Such a small uncertainty was neglected because it had a much
smaller impact on the resulting abundances than other sources of uncertainty
considered in Sections~\ref{sec-err1} and \ref{sec-err2}. 

%%%%%%%%%%%%%%%%%%%%%%%%%%%%%%%%%%%%%%%%%%%%%%%%% Different Methods
\begin{figure}
\centering
\includegraphics[width=\columnwidth]{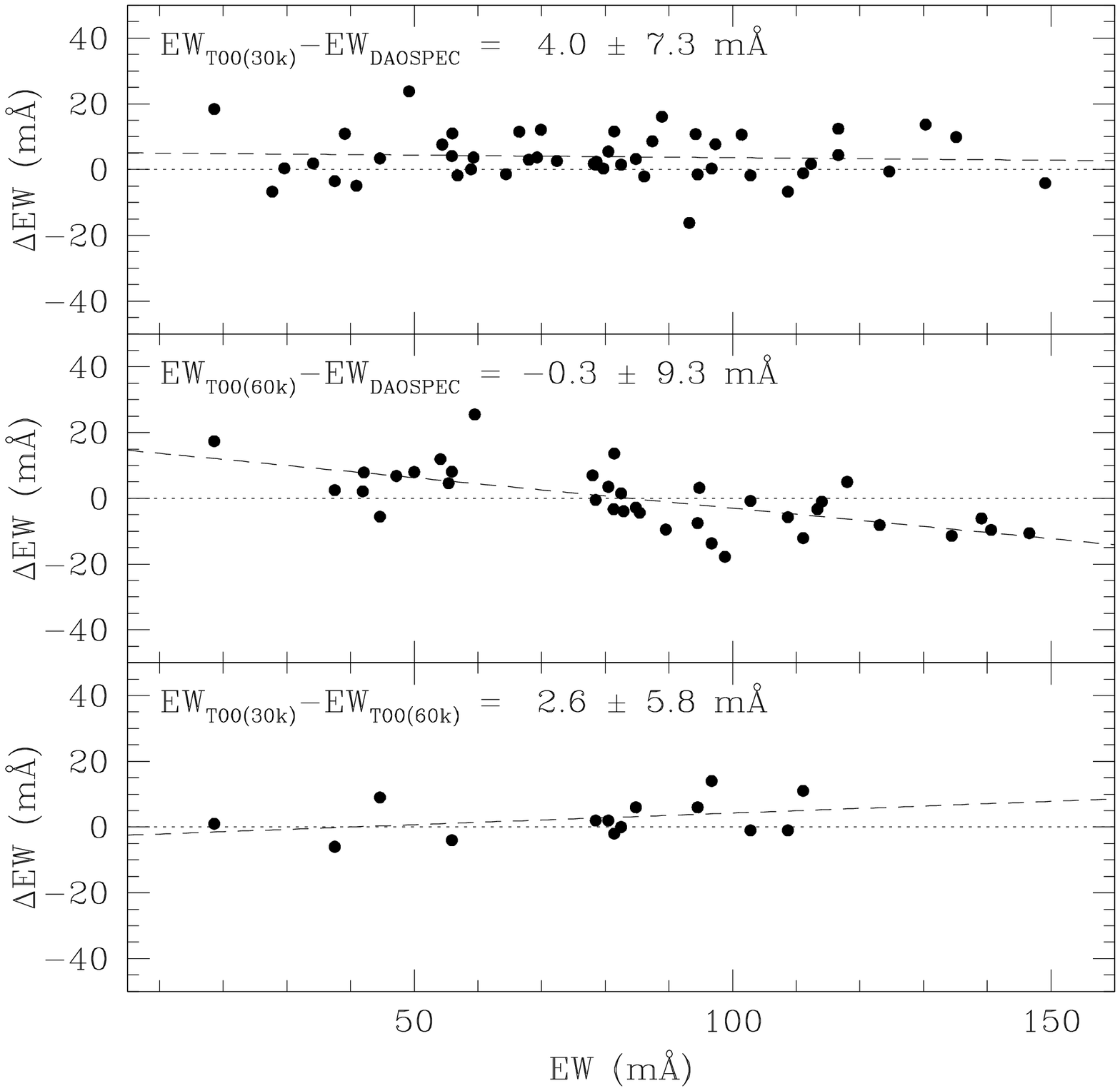}

\caption{Comparison of our EW measurements with \citet{taut00}, for star 141 in
M~67. The top panel shows the comparison of DAOSPEC EW with the R$\simeq$30000
set by \citet{taut00}, based on 48 lines in common. The middle panel show the
same comparison, but for the R$\simeq$60000 set by \citet{taut00}, based on 36
lines. The bottom panel shows the 15 lines in common between the R$\simeq$30000
and the R$\simeq$60000 EW sets, considering only the lines included in the two
upper panels.}

\label{taut}
\end{figure}
%%%%%%%%%%%%%%%%%%%%%%%%%%%%%%%%%%%%%%%%%%%%%%%%% Different Methods

\subsection{Comparison with Literature EW}
\label{sec-taut}

To our knowledge, only one of our target stars was studied before by
\citet{yong05} and \citet{taut00}, with a resolution and S/N similar to ours,
i.e. star 141 in M~67. While \citet{yong05} do not publish their EW
measurements, we can compare with the ones by \citet{taut00}. The authors
provided two sets of EW, the former derived from a spectrum with R$\simeq$30000,
and the latter from a spectrum with R$\simeq$60000. We have 48 lines in common
with the R$\simeq$30000 set and 36 with the R$\simeq$60000 set. 

Figure~\ref{taut} shows good agreement between our EWs and the R$\simeq$30000
set. We found just a sistematic offset of
EW$_{T00}$-EW$_{DAOSPEC}=4.0\pm7.3$~m\AA, which corresponds to a continuum
placement difference of about 1\% (see also Section~\ref{sec-cont}). A possible
trend with EW was visible when comparing our EWs with the R$\simeq$60000 set,
with no systematic offset ($\Delta EW = -0.3 \pm 9.3$~m\AA). On the one hand,
this means that our continuum placement agrees much better with the
R$\simeq$60000 continuum by \citet{taut00} than with the R$\simeq$30000 one. On
the other hand, we noted that a possible trend is also visible when comparing
the \citet{taut00} measurements at R$\simeq$30000 with those at R$\simeq$60000.
In conclusion, we considered our EW measurements in good agreement with the
\citet{taut00} ones, given the involved uncertainties (see also
Table~\ref{tab-141}).

%__________________________________________________________________

\section{Abundance Analysis}
\label{sec-abo}

\subsection{Best Model Search}
\label{sec-best}

A preliminary abundance determination was done using the photometric parameters
(Section~\ref{phot}), which allowed us to identify and remove those lines in our
list that gave systematically discrepant abundances. We found largely discrepant
Fe~I and Fe~II values when using the photometric parameters (by
$\simeq$0.5--0.9~dex), indicating that something was wrong with the photometric
gravities (Section~\ref{sec-phot}).

As a second step, we calculated Fe~I and Fe~II abundances for a set of
models with parameters extending more than 3$\sigma$ around the
photometric estimates of Table~\ref{pars}, i.e., about $\pm$300--500~K in
$T_{\rm{eff}}$, $\pm$0.3--0.6~dex in $\log g$ and $\pm$0.5~km~s$^{-1}$
in $v_t$, depending on the star. This large grid of calculated
abundances was used to refine our photometric estimate of the
atmospheric parameters. 

We chose the model that satisfied simultaneously (within the
uncertainties\footnote{We basically considered a slope consistent with zero when
the 3~$\sigma$ spread around the fit was larger than the maximum [Fe/H] difference
implied by the fitted slope at the extremes of the interval covered by  the
abscissae (be it $\chi_{ex}$, EW or $\lambda$).}) the
following conditions: {\em (i)} the abundance of Fe~I lines should not vary with
excitation potential $\chi_{ex}$; {\em (ii)} the abundance of Fe~I lines should
not vary significantly with EW, i.e., strong and weak lines should give the same
abundance\footnote{We decided not to use the \citet{mag84} effect, because we
prefer to have internally consistent abundances from each line, and the
difference between the two methods in our case appears small ($\Delta v_t <
0.2$~km~s$^{-1}$).}; {\em (iii)} the abundance of Fe~I lines should not differ
sigificantly from the abundance of Fe~II lines; {\em (iv)} the abundance of Fe~I
lines should not vary significantly with wavelength. 

%%%%%%%%%%%%%%%%%%%%%%%%%%%%%%%%%%%%%%%%%%%%%%%% M67 141
\begin{table}
\begin{minipage}[htb]{\columnwidth}
\caption{Literature comparison for star 141 in M~67.}             
\label{tab-141}      
\centering          
\renewcommand{\footnoterule}{}  
\begin{tabular}{l r r r }     
\hline\hline       
Parameter         & Here            & T00             & Y05     \\
\hline     	     
Resolution        & 30000           & 30--60000       & 28000   \\
S/N               & 70-100          & $\geq$100       & 30--100 \\
T$_eff$ (K)       & 4650            & 4730            & 4700    \\
log$g$ (dex)      & 2.8             & 2.4             & 2.3     \\
$v_t$ (km~s$^{-1}$) & 1.3           & 1.8             & 1.3     \\
\hline                  
$[$FeI/H$]$       &   0.06$\pm$0.01($\pm$0.10) & --0.01$\pm$0.02 & --0.01$\pm$0.02 \\ 
$[$FeII/H$]$      &   0.01$\pm$0.03($\pm$0.10) & --0.01$\pm$0.05 &   0.01$\pm$0.03 \\
$[$$\alpha$/Fe$]$ &   0.01$\pm$0.01($\pm$0.07) &   0.01$\pm$0.03 &   0.12$\pm$0.06\\
\hline				   			
$[$Al/Fe$]$       &   0.06$\pm$0.06($\pm$0.05) &   0.08$\pm$0.01 &   0.16$\pm$0.03 \\  
$[$Ba/Fe$]$       &   0.26$\pm$0.05($\pm$0.06) &   0.07$\pm$0.00 &   0.02$\pm$0.00 \\ 
$[$Ca/Fe$]$       & --0.13$\pm$0.02($\pm$0.03) &   0.09$\pm$0.05 &   0.09$\pm$0.02 \\  
$[$Co/Fe$]$       &   0.11$\pm$0.02($\pm$0.06) &   0.05$\pm$0.07 &   0.01$\pm$0.05 \\ 
$[$Cr/Fe$]$       &   0.01$\pm$0.03($\pm$0.06) &   0.12$\pm$0.05 &   ---  	 \\ 
$[$La/Fe$]$       &   0.06$\pm$0.05($\pm$0.09) & --0.06$\pm$0.00 &   0.13$\pm$0.03 \\ 
$[$Mg/Fe$]$       &   0.29$\pm$0.03($\pm$0.10) &   0.11$\pm$0.00 &   0.18$\pm$0.02 \\ 
$[$Na/Fe$]$       &   0.10$\pm$0.02($\pm$0.04) &   0.25$\pm$0.00 &   0.24$\pm$0.06 \\ 
$[$Nd/Fe$]$       &   0.01$\pm$0.29($\pm$0.10) &     ---         &   ---  	 \\ 
$[$Ni/Fe$]$       &   0.06$\pm$0.02($\pm$0.03) &   0.05$\pm$0.04 &   0.06$\pm$0.03 \\ 
$[$O/Fe$]$        & --0.05$\pm$0.09($\pm$0.08) &   0.04$\pm$0.00 &   0.10$\pm$0.04 \\ 
$[$Sc/Fe$]$       & --0.02$\pm$0.08($\pm$0.08) &   0.09$\pm$0.05 &   ---		\\ 
$[$Si/Fe$]$       &   0.09$\pm$0.02($\pm$0.08) &   0.11$\pm$0.04 &   0.11$\pm$0.04 \\ 
$[$TiI/Fe$]$      & --0.07$\pm$0.02($\pm$0.07) &   0.04$\pm$0.05 &   0.05$\pm$0.03 \\
$[$TiII/Fe$]$     & --0.07$\pm$0.08($\pm$0.06) &   0.11$\pm$0.07 &   --- \\
$[$V/Fe$]$        &   0.13$\pm$0.04($\pm$0.14) &   0.09$\pm$0.03 &   --- \\
$[$Y/Fe$]$        & --0.04$\pm$0.02($\pm$0.12) & --0.12$\pm$0.08 &   --- \\
\hline \hline 
\end{tabular}
\end{minipage}
\end{table}
%%%%%%%%%%%%%%%%%%%%%%%%%%%%%%%%%%%%%%%%%%%%%%%% M67 141

Using the spreads of the Fe~I and Fe~II abundances of various lines, and the
uncertainties in the slopes of the above conditions, we estimated the typical
1$\sigma$ uncertainties on the spectroscopic parameters: about $\pm$100~K on
$T_{\rm{eff}}$, $\pm$0.2~dex on $\log g$ and $\pm$0.1~km~s$^{-1}$ on $v_t$.

The resulting spectroscopic parameters (Table~\ref{pars}) were always in good
agreement with the photometric ones, within the quoted uncertainties, with a
tendency of the spectroscopic gravities to be systematically higher by
0.3--0.5~dex (see above). However, these differences are always easily
accomodated within the uncertainty ranges (due to photometric errors,
uncertainties in distance moduli and reddenings, and bolometric corrections).  

\subsection{Abundance Calculations}
\label{abo-calc}

Abundance calculations and spectral synthesis (for oxygen) were done using the
last updated version of the abundance calculation code originally described by
\citet{spi67}. We used the model atmospheres by \citet{edv93}. We also made use
of ABOMAN, a tool developed by E. Rossetti at the INAF, Bologna Observatory,
Italy, that allows for the semi-automatic processing of several objects, using
the above abundance calculation codes. ABOMAN performs automatically all the
steps needed to choose the best model, and provides all the graphical tools to
analyze the results. 

%%%%%%%%%%%%%%%%%%%%%%%%%%%%%%%%%%%%%%%%%%%%%% Abundances
\begin{table*}[!t]
\begin{minipage}[t]{\textwidth}
\caption{Abundance Ratios for sigle cluster stars, with their 
{\em internal} uncertainties (Section~\ref{sec-err1}). For {\em external}
uncertainties see Table~\ref{errpar}}      
\label{abotab}  
\centering          
\begin{tabular}{l|c c c|c c c|c c c}  
\hline \hline    
& \multicolumn{3}{c|}{Cr~110} & \multicolumn{3}{c|}{NGC~2099 (M~37)} & \multicolumn{3}{c}{NGC~2420} \\        	  					    
Ratio         & Star 2108       & Star 2129       & Star 3144       & Star 067        & Star 148        & Star 508        & Star 041        & Star 076        & Star 174    \\                
\hline
$[$FeI/H$]$       &   0.02$\pm$0.01 &   0.05$\pm$0.01 &   0.02$\pm$0.01 &   0.01$\pm$0.01 & --0.03$\pm$0.01 &   0.07$\pm$0.01 & --0.07$\pm$0.01 & --0.06$\pm$0.01 & --0.03$\pm$0.01 \\ 
$[$FeII/H$]$      &   0.00$\pm$0.06 & --0.04$\pm$0.08 &   0.00$\pm$0.06 & --0.01$\pm$0.02 & --0.07$\pm$0.04 &   0.05$\pm$0.06 & --0.09$\pm$0.06 &   0.00$\pm$0.02 & --0.07$\pm$0.04 \\
$[$$\alpha$/Fe$]$ & --0.02$\pm$0.02 &   0.01$\pm$0.02 &   0.04$\pm$0.02 & --0.04$\pm$0.01 &   0.00$\pm$0.01 &   0.03$\pm$0.01 & --0.03$\pm$0.01 & --0.03$\pm$0.01 &   0.03$\pm$0.02 \\
\hline
$[$Al/Fe$]$   & --0.02$\pm$0.04 & --0.06$\pm$0.08 & --0.03$\pm$0.06 & --0.06$\pm$0.08 & --0.10$\pm$0.07 & --0.04$\pm$0.07 & --0.10$\pm$0.04 & --0.10$\pm$0.05 & --0.09$\pm$0.05 \\
$[$Ba/Fe$]$   &   0.48$\pm$0.02 &   0.49$\pm$0.04 &   0.52$\pm$0.06 &	0.60$\pm$0.05 &   0.55$\pm$0.04 &   0.58$\pm$0.04 &   0.54$\pm$0.04 &	0.58$\pm$0.05 &   0.65$\pm$0.02 \\
$[$Ca/Fe$]$   & --0.08$\pm$0.04 & --0.04$\pm$0.06 & --0.09$\pm$0.07 & --0.06$\pm$0.05 & --0.08$\pm$0.04 & --0.09$\pm$0.04 & --0.08$\pm$0.03 & --0.11$\pm$0.04 & --0.08$\pm$0.05 \\
$[$Co/Fe$]$   &   0.13$\pm$0.05 & --0.08$\pm$0.04 &   0.01$\pm$0.04 & --0.04$\pm$0.02 & --0.04$\pm$0.02 & --0.05$\pm$0.02 &   0.10$\pm$0.05 & --0.02$\pm$0.03 &   0.05$\pm$0.03 \\ 
$[$Cr/Fe$]$   &   0.00$\pm$0.04 &   0.06$\pm$0.06 &   0.03$\pm$0.08 & --0.01$\pm$0.06 &   0.01$\pm$0.05 &   0.05$\pm$0.05 & --0.02$\pm$0.03 & --0.09$\pm$0.06 & --0.02$\pm$0.05 \\
$[$La/Fe$]$   &   0.11$\pm$0.09 &   0.12$\pm$0.03 & --0.15$\pm$0.04 &	0.08$\pm$0.02 &   0.16$\pm$0.08 &   0.13$\pm$0.08 &   0.23$\pm$0.01 &	0.12$\pm$0.09 &   0.07$\pm$0.13 \\
$[$Mg/Fe$]$   &   0.19$\pm$0.07 &   0.01$\pm$0.14 &   0.16$\pm$0.08 &	0.28$\pm$0.05 &   0.26$\pm$0.04 &   0.27$\pm$0.03 &   0.09$\pm$0.05 &	0.10$\pm$0.09 &   0.14$\pm$0.08 \\
$[$Na/Fe$]$   & --0.08$\pm$0.02 & --0.06$\pm$0.08 &   0.00$\pm$0.03 &	0.08$\pm$0.05 &   0.10$\pm$0.04 &   0.05$\pm$0.08 & --0.13$\pm$0.09 & --0.04$\pm$0.06 &   0.02$\pm$0.07 \\
$[$Nd/Fe$]$   &   0.05$\pm$0.16 &   0.29$\pm$0.13 &   0.51$\pm$0.28 &	0.23$\pm$0.29 &   0.26$\pm$0.25 &   0.33$\pm$0.32 &   0.12$\pm$0.16 &	0.40$\pm$0.28 &   0.18$\pm$N.A. \\
$[$Ni/Fe$]$   & --0.02$\pm$0.02 & --0.01$\pm$0.02 & --0.04$\pm$0.03 & --0.04$\pm$0.03 & --0.02$\pm$0.02 & --0.01$\pm$0.02 &   0.03$\pm$0.02 & --0.01$\pm$0.02 & --0.01$\pm$0.02 \\
$[$O/Fe$]$    &   0.08$\pm$0.08 & --0.07$\pm$0.12 &   0.02$\pm$0.14 &   0.12$\pm$0.07 &   0.25$\pm$0.15 &   0.22$\pm$0.13 & --0.01$\pm$0.16 &   0.24$\pm$0.14 &   0.39$\pm$0.33 \\
$[$Sc/Fe$]$   &   0.11$\pm$0.05 & --0.07$\pm$0.06 & --0.14$\pm$0.10 & --0.05$\pm$0.05 & --0.01$\pm$0.07 & --0.02$\pm$0.12 &   0.07$\pm$0.01 &	0.11$\pm$0.07 &   0.16$\pm$0.06 \\
$[$Si/Fe$]$   &   0.02$\pm$0.03 &   0.04$\pm$0.02 &   0.06$\pm$0.03 &	0.07$\pm$0.03 &   0.11$\pm$0.02 &   0.06$\pm$0.03 &   0.05$\pm$0.02 &	0.05$\pm$0.02 &   0.02$\pm$0.03 \\
$[$TiI/Fe$]$  &   0.01$\pm$0.03 &   0.00$\pm$0.03 &   0.01$\pm$0.04 & --0.10$\pm$0.02 & --0.10$\pm$0.02 & --0.07$\pm$0.02 & --0.06$\pm$0.02 & --0.09$\pm$0.02 &   0.06$\pm$0.03 \\
$[$TiII/Fe$]$ &   0.01$\pm$0.13 & --0.04$\pm$0.07 &   0.11$\pm$0.05 & --0.01$\pm$0.06 &   0.00$\pm$0.05 &   0.05$\pm$0.08 & --0.04$\pm$0.23 &	0.07$\pm$0.09 &   0.12$\pm$0.07 \\
$[$V/Fe$]$    &   0.28$\pm$0.06 & --0.03$\pm$0.05 & --0.09$\pm$0.07 & --0.09$\pm$0.02 & --0.08$\pm$0.04 & --0.03$\pm$0.04 &   0.11$\pm$0.06 & --0.08$\pm$0.05 &   0.13$\pm$0.04 \\
$[$Y/Fe$]$    &   0.02$\pm$0.12 & --0.17$\pm$0.08 & --0.03$\pm$0.18 & --0.03$\pm$0.23 &   0.09$\pm$0.09 &   0.08$\pm$0.20 & --0.14$\pm$0.14 &	0.06$\pm$0.02 &   0.02$\pm$0.09 \\
\hline \hline    
& \multicolumn{3}{c|}{NGC~2682 (M~67)} & \multicolumn{3}{c|}{NGC~7789} \\ 		     	             	  					    
Ratio         & Star 141        & Star 223        & Star 286  & Star 5237       & Star 7840	  & Star 8556	          \\                
\hline 
$[$FeI/H$]$       &   0.06$\pm$0.01 &   0.04$\pm$0.01 &   0.06$\pm$0.01 & --0.01$\pm$0.01 &   0.03$\pm$0.01 &   0.12$\pm$0.01 \\  
$[$FeII/H$]$      &   0.01$\pm$0.03 &   0.02$\pm$0.05 & --0.05$\pm$0.04 & --0.06$\pm$0.05 & --0.03$\pm$0.07 &   0.07$\pm$0.06 \\  
$[$$\alpha$/Fe$]$ &   0.01$\pm$0.01 & --0.03$\pm$0.01 & --0.05$\pm$0.01 & --0.02$\pm$0.02 & --0.02$\pm$0.02 & --0.08$\pm$0.02 \\
\hline 
$[$Al/Fe$]$   &   0.06$\pm$0.06 &   0.02$\pm$0.05 &   0.03$\pm$0.05 & --0.01$\pm$0.10 &   0.05$\pm$0.11 & --0.13$\pm$0.10 \\  
$[$Ba/Fe$]$   &   0.26$\pm$0.05 &   0.27$\pm$0.05 &   0.23$\pm$0.05 &	0.49$\pm$0.06 &   0.36$\pm$0.07 &   0.46$\pm$0.03 \\  
$[$Ca/Fe$]$   & --0.13$\pm$0.02 & --0.20$\pm$0.03 & --0.15$\pm$0.02 & --0.10$\pm$0.04 & --0.12$\pm$0.07 & --0.24$\pm$0.03 \\  
$[$Co/Fe$]$   &   0.11$\pm$0.02 &   0.05$\pm$0.04 &   0.17$\pm$0.04 & --0.02$\pm$0.04 & --0.03$\pm$0.02 & --0.01$\pm$0.03 \\  
$[$Cr/Fe$]$   &   0.01$\pm$0.03 & --0.01$\pm$0.04 &   0.06$\pm$0.05 &	0.05$\pm$0.04 & --0.02$\pm$0.03 &   0.16$\pm$0.45 \\  
$[$La/Fe$]$   &   0.06$\pm$0.05 & --0.06$\pm$0.06 &   0.07$\pm$0.03 &	0.14$\pm$0.06 &   0.06$\pm$0.07 &   0.14$\pm$0.15 \\  
$[$Mg/Fe$]$   &   0.29$\pm$0.03 &   0.20$\pm$0.08 &   0.23$\pm$0.05 &	0.26$\pm$0.07 &   0.24$\pm$0.04 &   0.13$\pm$0.06 \\  
$[$Na/Fe$]$   &   0.10$\pm$0.02 & --0.02$\pm$0.04 &   0.22$\pm$0.08 &	0.04$\pm$0.01 &   0.09$\pm$0.08 & --0.15$\pm$0.01 \\  
$[$Nd/Fe$]$   &   0.01$\pm$0.29 &   0.10$\pm$0.25 &   0.10$\pm$0.21 &	0.21$\pm$0.38 & --0.17$\pm$0.27 &   0.35$\pm$0.20 \\  
$[$Ni/Fe$]$   &   0.06$\pm$0.02 &   0.04$\pm$0.02 &   0.04$\pm$0.02 & --0.02$\pm$0.02 &   0.00$\pm$0.02 & --0.02$\pm$0.03 \\  
$[$O/Fe$]$    & --0.05$\pm$0.09 &   0.09$\pm$0.11 &   0.13$\pm$0.10 &	0.21$\pm$0.15 &   0.13$\pm$0.10 &   0.23$\pm$0.20 \\  
$[$Sc/Fe$]$   & --0.02$\pm$0.08 & --0.05$\pm$0.04 &   0.01$\pm$0.05 &	0.10$\pm$0.09 &   0.06$\pm$0.10 &   0.08$\pm$0.12 \\  
$[$Si/Fe$]$   &   0.09$\pm$0.02 &   0.12$\pm$0.02 &   0.06$\pm$0.02 & --0.01$\pm$0.02 & --0.02$\pm$0.03 &   0.03$\pm$0.04 \\  
$[$TiI/Fe$]$  & --0.07$\pm$0.02 & --0.11$\pm$0.02 &   0.00$\pm$0.02 & --0.02$\pm$0.03 & --0.03$\pm$0.03 & --0.14$\pm$0.04 \\  
$[$TiII/Fe$]$ & --0.07$\pm$0.08 &   0.02$\pm$0.06 & --0.06$\pm$0.13 & --0.07$\pm$0.29 & --0.19$\pm$0.17 &   0.14$\pm$0.05 \\  
$[$V/Fe$]$    &   0.13$\pm$0.04 &   0.06$\pm$0.04 &   0.25$\pm$0.04 &	0.05$\pm$0.04 & --0.01$\pm$0.04 & --0.16$\pm$0.06 \\  
$[$Y/Fe$]$    & --0.04$\pm$0.02 & --0.06$\pm$0.05 &   0.00$\pm$0.03 &	0.22$\pm$N.A. & --0.01$\pm$N.A. &   0.07$\pm$0.18 \\						 
\hline \hline          	  					    
\end{tabular}	   
\end{minipage}	   
\end{table*}	   
%%%%%%%%%%%%%%%%%%%%%%%%%%%%%%%%%%%%%%%%%%%%%% Abundances

When the best model was found for each star, abundances and abundance ratios of
all the species of interest were calculated, by averaging the results for each
line of that element (Table~\ref{abotab}). Abundance ratios were always computed
with respect to Fe~I. Cluster averages were computed as weighted averages of the
results for each star in the cluster and, if necessary, of different ionization
stages of each element (Table~\ref{tab-cluster}). In all Tables, [$\alpha$/Fe]
is the weighted average of all $\alpha$-elements abundances.

We can compare our results for star 141 in M~67 (see also
Section~\ref{sec-taut}) with the ones by \citet{taut00} and \citet{yong05}. We
find a general agreement for both atmospheric parameter and abundance ratios,
with the only significant exception of barium (but see the discussion in
Section~\ref{heavy-abo}), calcium (discussed in Section~\ref{alfa-abo}) and
titanium. For all the discussed elements our results, as discussed in
Section~\ref{sec-disc}, are in broad agreement with the whole body of
high resolution abundances for OC, so we consider our results sound.

%%%%%%%%%%%%%%%%%%%%%%%%%%%%%%%%%%%%%%%%%%%%%%%%%% Cluster Averages
\begin{table*}
\begin{minipage}[t]{\textwidth}

\caption{Average cluster abundances, obtained with the weighted average of the
single stars abundances.}             

\label{tab-cluster} 
\centering          
\renewcommand{\footnoterule}{}  
\begin{tabular}{l c c c c c}     
\hline\hline 
Ratio           &  Cr~110         & NGC~2099        & NGC~2420        & M~67            & NGC~7789 \\ 
                & 	          &		    &	              &                 & \\
\hline
$[$Fe/H$]$      &  +0.03$\pm$0.02($\pm$0.10) &  +0.01$\pm$0.05($\pm$0.10) & --0.05$\pm$0.03($\pm$0.10) &  +0.05$\pm$0.02($\pm$0.10) &  +0.04$\pm$0.07($\pm$0.10) \\
$[\alpha$/Fe$]$ &  +0.01$\pm$0.03($\pm$0.07) &   0.00$\pm$0.04($\pm$0.07) & --0.02$\pm$0.03($\pm$0.07) & --0.02$\pm$0.03($\pm$0.07) & --0.04$\pm$0.03($\pm$0.07) \\
\hline  			 	    		        	 		   	      				   		     		
$[$Al/Fe$]$     & --0.04$\pm$0.02($\pm$0.05) & --0.06$\pm$0.03($\pm$0.05) & --0.10$\pm$0.01($\pm$0.05) &  +0.03$\pm$0.02($\pm$0.05) & --0.03$\pm$0.09($\pm$0.05) \\
$[$Ba/Fe$]$     &  +0.49$\pm$0.02($\pm$0.06) &  +0.57$\pm$0.02($\pm$0.06) &  +0.62$\pm$0.07($\pm$0.06) &  +0.25$\pm$0.02($\pm$0.06) &  +0.45$\pm$0.05($\pm$0.06) \\
$[$Ca/Fe$]$     & --0.07$\pm$0.02($\pm$0.03) & --0.08$\pm$0.01($\pm$0.03) & --0.09$\pm$0.02($\pm$0.03) & --0.16$\pm$0.03($\pm$0.03) & --0.18$\pm$0.09($\pm$0.03) \\ 
$[$Co/Fe$]$     &  +0.01$\pm$0.10($\pm$0.06) & --0.04$\pm$0.01($\pm$0.06) &  +0.03$\pm$0.06($\pm$0.06) &  +0.08$\pm$0.06($\pm$0.06) & --0.02$\pm$0.01($\pm$0.06) \\ 
$[$Cr/Fe$]$     &  +0.02$\pm$0.03($\pm$0.06) &  +0.02$\pm$0.03($\pm$0.06) & --0.03$\pm$0.03($\pm$0.06) &  +0.01$\pm$0.03($\pm$0.06) &  +0.01$\pm$0.05($\pm$0.06) \\ 
$[$La/Fe$]$ 	&  +0.03$\pm$0.18($\pm$0.09) &  +0.09$\pm$0.05($\pm$0.09) &  +0.23$\pm$0.09($\pm$0.09) &  +0.05$\pm$0.06($\pm$0.09) &  +0.11$\pm$0.05($\pm$0.09) \\ 
$[$Mg/Fe$]$ 	&  +0.16$\pm$0.07($\pm$0.10) &  +0.27$\pm$0.01($\pm$0.10) &  +0.09$\pm$0.06($\pm$0.10) &  +0.27$\pm$0.04($\pm$0.10) &  +0.22$\pm$0.07($\pm$0.10) \\ 
$[$Na/Fe$]$ 	& --0.06$\pm$0.05($\pm$0.04) &  +0.09$\pm$0.02($\pm$0.04) & --0.04$\pm$0.07($\pm$0.04) &  +0.08$\pm$0.09($\pm$0.04) & --0.05$\pm$0.13($\pm$0.04) \\ 
$[$Nd/Fe$]$ 	&  +0.23$\pm$0.20($\pm$0.10) &  +0.27$\pm$0.05($\pm$0.10) &  +0.19$\pm$0.17($\pm$0.10) &  +0.08$\pm$0.05($\pm$0.10) &  +0.17$\pm$0.30($\pm$0.10) \\ 
$[$Ni/Fe$]$ 	& --0.02$\pm$0.01($\pm$0.03) & --0.02$\pm$0.01($\pm$0.03) &  +0.00$\pm$0.02($\pm$0.03) &  +0.05$\pm$0.01($\pm$0.03) & --0.01$\pm$0.01($\pm$0.03) \\ 
$[$O/Fe$]$  	&  +0.03$\pm$0.08($\pm$0.08) &  +0.15$\pm$0.08($\pm$0.08) &  +0.15$\pm$0.18($\pm$0.08) &  +0.04$\pm$0.10($\pm$0.08) &  +0.16$\pm$0.06($\pm$0.08) \\ 
$[$Sc/Fe$]$ 	&  +0.01$\pm$0.13($\pm$0.08) & --0.03$\pm$0.03($\pm$0.08) &  +0.07$\pm$0.05($\pm$0.08) & --0.03$\pm$0.04($\pm$0.08) &  +0.08$\pm$0.02($\pm$0.08) \\ 
$[$Si/Fe$]$ 	&  +0.04$\pm$0.02($\pm$0.07) &  +0.09$\pm$0.03($\pm$0.07) &  +0.04$\pm$0.01($\pm$0.07) &  +0.10$\pm$0.02($\pm$0.07) & --0.01$\pm$0.02($\pm$0.07) \\ 
$[$Ti/Fe$]$ 	&  +0.02$\pm$0.04($\pm$0.06) & --0.08$\pm$0.04($\pm$0.06) & --0.04$\pm$0.08($\pm$0.06) & --0.04$\pm$0.06($\pm$0.06) & --0.03$\pm$0.09($\pm$0.06) \\ 
$[$V/Fe$]$  	&  +0.05$\pm$0.19($\pm$0.14) & --0.08$\pm$0.03($\pm$0.14) & --0.05$\pm$0.13($\pm$0.14) &  +0.15$\pm$0.13($\pm$0.14) & --0.01$\pm$0.09($\pm$0.14) \\ 
$[$Y/Fe$]$  	& --0.10$\pm$0.12($\pm$0.12) &  +0.07$\pm$0.06($\pm$0.12) &  +0.05$\pm$0.08($\pm$0.12) & --0.05$\pm$0.04($\pm$0.12) &  +0.08$\pm$0.09($\pm$0.12) \\ 
\hline             	  	 
\end{tabular}	   		 
\end{minipage}	   
\end{table*}	   
%%%%%%%%%%%%%%%%%%%%%%%%%%%%%%%%%%%%%%%%%%%%%%%%%% Cluster Averages

\subsection{Internal Abundance Uncertainties}
\label{sec-err1}

Random uncertainties in the EW measurement process and in the log$gf$
determinations were taken into account by averaging the abundances
determinations obtained for different lines and using $\sigma /
\sqrt(n_{lines})$ as the final {\em internal} error. These are reported in
Table~\ref{abotab}, and they are of the order of $\sim$0.01~dex  for Fe~I which
had the highest number of lines, and could reach up to $\sim$0.2--0.3~dex for
elements relying on a handful of lines such as Nd, for example. Additional
systematic (from line to line) and random (from star to star) uncertainties in
EW measurements, due to continuum placement, had a negligible impact on the
final abundances in our specific case (Section~\ref{sec-cont}). 

For the spectral synthesis of oxygen, the uncertainty of the fit was computed by
bracketing the best fitting synthetic spectrum with two spectra with altered
abundance. The bracketing spectra were chosen to overlap the 1$\sigma$
poissonian uncertainty on the observed spectrum. The abundance difference of the
two bracketing spectra with the best fit were averaged together to produce the
estimated uncertainty, reported in Table~\ref{abotab} (and in our solar
analysis, Table~\ref{sun}).

\subsection{Uncertainties due to the Choice of Stellar Parameters}
\label{sec-err2}

The choice of stellar parameters implies systematic (from line to line) and random
(from star to star) uncertainties on the final abundances. Usually, to estimate
the impact of the stellar parameter choice on the derived abundances, each
parameter is altered by its estimated uncertainty and the resulting abundance
differences with respect to the best model parameter abundance set are summed in
quadrature to obtain a global uncertainty. We applied this method to our coolest
(508 in NGC~2099) and warmest star (2129 in Cr~110) and obtained for
the various abundance ratios uncertainties ranging from 0.05 to 0.45~dex, with a
typical value around $\sim$0.10~dex.

However, as noted by \citet{cay04}, this method produces a very conservative
estimate of the uncertainty, because it neglects the natural correlation among
stellar parameters occurring when the so called "spectroscopic method"
(Section~\ref{sec-best}) is adopted. Covariance terms should therefore be
included to properly take into account such dependencies among the parameters
\citep[see][for a detailed treatment of the problem]{mcw95}. The practical
method proposed by \citet{cay04} assumes that T$_{\rm{eff}}$ largely dominates
the abundance results and, therefore, T$_{\rm{eff}}$ has to be varied by its
estimated uncertainty ($\simeq$100~K in our case, Section~\ref{sec-best}). A new
"second best" model has to be identified with the new value of T$_{\rm{eff}}$ by
varying $v_t$ and $\log g$ accordingly, to minimize as much as possible the
slopes of the relations described in Section~\ref{sec-best}. This method
naturally and properly takes into account both the main terms of the error
budget {\em and} the appropriate covariance terms.

We therefore altered the temperature of our hottest and coolest stars (see
above) by +100~K and by --100~K. We re-optimized all the parameters and
re-calculated all the abundance ratios. The final uncertainties are the average
of the uncertainties calculated with the higher and lower temperature and are
shown in Table~\ref{errpar}. The average between the uncertainties of these two
cases is taken as a reliable estimate for the impact of the choice of stellar
parameters on our abundance ratios. We added these {\em external} uncertainties
between parenthesis after each abundance ratio and we summed them in quadrature
with the {\em internal} errors to produce the errobars in each Figure. 

%%%%%%%%%%%%%%%%%%%%%%%%%%%%%%%%%%%%%%%%%%%%%%%%%%%%%% Error Analysis
\begin{table}
\begin{minipage}[t]{\columnwidth}
\caption{Uncertanties due to the choice of stellar parameters.}             
\label{errpar}      
\centering          
\renewcommand{\footnoterule}{}  
\begin{tabular}{l c c c}     
\hline\hline       
Ratio & NGC2099-508  & Cr110-2129 & Average \\
      & (T$=$4500~K) & (T$=$4950~K) & \\
%        & & & \\
\hline      
$[$Al/Fe$]$   & $\pm$0.04 & $\pm$0.07 & $\pm$0.05 \\
$[$Ba/Fe$]$   & $\pm$0.04 & $\pm$0.08 & $\pm$0.06 \\
$[$Ca/Fe$]$   & $\pm$0.03 & $\pm$0.03 & $\pm$0.03 \\
$[$Co/Fe$]$   & $\pm$0.07 & $\pm$0.05 & $\pm$0.06 \\
$[$Cr/Fe$]$   & $\pm$0.09 & $\pm$0.03 & $\pm$0.06 \\
$[$Fe/H$]$    & $\pm$0.16 & $\pm$0.04 & $\pm$0.10 \\
$[$La/Fe$]$   & $\pm$0.10 & $\pm$0.07 & $\pm$0.09 \\
$[$Mg/Fe$]$   & $\pm$0.11 & $\pm$0.09 & $\pm$0.10 \\
$[$Na/Fe$]$   & $\pm$0.03 & $\pm$0.06 & $\pm$0.04 \\
$[$Nd/Fe$]$   & $\pm$0.09 & $\pm$0.12 & $\pm$0.10 \\
$[$Ni/Fe$]$   & $\pm$0.02 & $\pm$0.03 & $\pm$0.03 \\	    
$[$O/Fe$]$    & $\pm$0.10 & $\pm$0.07 & $\pm$0.08 \\
$[$Sc/Fe$]$   & $\pm$0.07 & $\pm$0.09 & $\pm$0.08 \\
$[$Si/Fe$]$   & $\pm$0.12 & $\pm$0.03 & $\pm$0.07 \\
$[$Ti/Fe$]$   & $\pm$0.07 & $\pm$0.06 & $\pm$0.06 \\
$[$V/Fe$]$    & $\pm$0.21 & $\pm$0.07 & $\pm$0.14 \\
$[$Y/Fe$]$    & $\pm$0.15 & $\pm$0.08 & $\pm$0.12 \\
		                
\hline            
\end{tabular}							     
\end{minipage}	   
\end{table}	   
%%%%%%%%%%%%%%%%%%%%%%%%%%%%%%%%%%%%%%%%%%%%%%%%%%%%%% Error Analysis

\subsection{Other Sources of Uncertainty}

The following additional sources of systematic uncertainties are not explicitly
discussed here, but should be taken into account when comparing our abundance
estimates with other works in the literature:

\begin{itemize}

\item{systematic uncertainties due to the choice of the solar reference
abundances, which are not discussed here. Our abundances can be reported to any
solar reference abundance with the information in Table~\ref{sun};}

\item{uncertainties due to the choice of $\log gf$ values, which can be estimated
by comparing our $\log gf$ values with other literature values (see
Section~\ref{ato-alpha});}

\item{uncertainties in the whole analysis due to more sophisticated effects
such as, NLTE, HFS, isotope ratios, that are difficult to evaluate in some cases
(these are mentioned whenever known or relevant in Sections~\ref{sec-li} and
\ref{sec-disc}; }

\item{small additional uncertainties due to the particular choice of atmospheric
models and of the abundance calculation code.}

\end{itemize}

%%%%%%%%%%%%%%%%%%%%%%%%%%%%%%%%%%%%%%%%%%%%%%%%%%%%%% Sun Analysis
\begin{table}
\begin{minipage}[t]{\columnwidth}

\caption{Solar abundance values.}             

\label{sun}      
\centering          
\renewcommand{\footnoterule}{}  
\begin{tabular}{l c c c c c}     
\hline\hline       
%HJD & $E$ & Method\#2 & \multicolumn{4}{c}{Method\#3}\\ 
Element & [X/H]$_{derived}$ & log$\epsilon_{GS96}$ & log$\epsilon_{A05}$ \\
%        & & & \\
\hline      
Al      & --0.34$\pm$N.A.($\pm$0.05) & 6.47$\pm$0.07 & 6.37$\pm$0.07 \\
Ba      &  +0.34$\pm$0.02($\pm$0.06) & 2.13$\pm$0.05 & 2.17$\pm$0.07 \\
Ca	& --0.09$\pm$0.03($\pm$0.03) & 6.36$\pm$0.02 & 6.31$\pm$0.04 \\
Co	&  +0.05$\pm$0.02($\pm$0.06) & 4.92$\pm$0.04 & 4.92$\pm$0.08 \\
Cr	&  +0.02$\pm$0.05($\pm$0.06) & 5.67$\pm$0.03 & 5.64$\pm$0.10 \\
FeI     &  +0.01$\pm$0.01($\pm$0.10) & 7.50$\pm$0.04 & 7.45$\pm$0.05 \\
FeII    & --0.03$\pm$0.02($\pm$0.10) & 7.50$\pm$0.04 & 7.45$\pm$0.05 \\
La      &   ---                      & 1.17$\pm$0.07 & 1.13$\pm$0.05 \\
Mg	&   ---                      & 7.58$\pm$0.05 & 7.53$\pm$0.09 \\
Na      & --0.10$\pm$0.01($\pm$0.04) & 6.33$\pm$0.03 & 6.17$\pm$0.04 \\
Nd      &   ---                      & 1.42$\pm$0.06 & 1.45$\pm$0.05 \\
Ni	&  +0.01$\pm$0.01($\pm$0.03) & 6.25$\pm$0.01 & 6.23$\pm$0.04 \\     
O       &  -0.01$\pm$0.07($\pm$0.08) & 8.87$\pm$0.07 & 8.66$\pm$0.05 \\
Sc	&  +0.06$\pm$0.02($\pm$0.08) & 3.17$\pm$0.10 & 3.05$\pm$0.08 \\
Si	& --0.08$\pm$0.01($\pm$0.07) & 7.55$\pm$0.05 & 7.51$\pm$0.04 \\
Ti	&  +0.03$\pm$0.05($\pm$0.06) & 5.02$\pm$0.06 & 4.90$\pm$0.06 \\
V 	& --0.14$\pm$0.02($\pm$0.14) & 4.00$\pm$0.02 & 4.00$\pm$0.02 \\
Y       &   ---                      & 2.24$\pm$0.03 & 2.21$\pm$0.02 \\
\hline             	  
\end{tabular}	   
\end{minipage}	   
\end{table}	   
%%%%%%%%%%%%%%%%%%%%%%%%%%%%%%%%%%%%%%%%%%%%%%%%%%%%%% Sun Analysis

\subsection{The Sun}
\label{sec-sun}

To test the whole abundance determination procedure, including EW
measurement, choice of lines and atomic parameters, and the uncertainties
determination criteria, we performed an abundance analysis of the Sun, and
checked that we obtain solar values for all elements, within the
uncertainties. We used the solar spectrum from the ESO spectrograph HARPS, in
La Silla, Chile, obtained by observing
Ganymede\footnote{http://www.ls.eso.org/lasilla/sciops/3p6/harps/monitoring/sun.html}.
The spectral resolution, R$\simeq$45000, is comparable to ours, while the
S/N$\simeq$350 is much higher.

To measure EWs, we used DAOSPEC and the same linelist used for our programme stars.
We then compared our solar EWs with two different literature sets, the first by
\citet{moore} and the second by \citet{rut84}. The median difference between
our EW and the ones by Moore is  EW$_{DAO}$--EW$_{Moore}$$=$0.9~m\AA\  (with an
inter-quartile range of $\pm$2.7~m\AA), based on 225 lines in common, while the one
with the \citet{rut84} measurements is EW$_{DAO}$--EW$_{RZ84}$$=$--0.5~m\AA\  (with
an inter-quartile range of $\pm$2.1~m\AA), based on 112 lines in common. For
completeness, we note that EW$_{Moore}$--EW$_{RZ84}$$=$1.5~m\AA\  (with an
inter-quartile range of $\pm$2.8~m\AA), based on 390 lines in common. We are
therefore satisfied with our solar EW measurements.

We then performed our abundance analysis (as in Section~\ref{abo-calc}) by
exploring the following atmospheric parameters ranges:  
$T_{\rm{eff,\odot}}$$=$5700--5800, in steps of 50~K; $\log g_\odot$$=$4.3--4.5,
in steps of 0.1~dex; and $v_t$$=$0.5--1.5, in steps of 0.1~km~s$^{-1}$. The
resulting best model has $T_{\rm{eff,\odot}}$$=$5750~K; $\log
g_\odot$$=$4.4~dex; and $v_t$$=$0.8~km~s$^{-1}$, in good agreement with the
accepted values \citep{iau}. Our adopted reference solar abundances
\citep{gre96} are shown in Table~\ref{sun}, along with the abundance ratios
derived as described. As can be seen, all the derived abundance ratios are
compatible with zero, within the uncertainties, with the exception of Al and Ba.
For Al, only one ($\lambda$$=$6698\AA) very weak (EW$=$18~m\AA) line could be
measured in the solar spectrum, while we used about 8 lines for the analysis of
our red clump giants. The lines at 6696 amd 6698\AA\  are known to give
tendentially lower values than the other Al lines \citep{red03,gra01}, so we do
not worry too much that our lone 6698\AA\footnote{Due to the shorter spectral
range of the solar spectrum and to a spectral defect, we could only measure one
Al line in the Sun.}  solar line gives a low [Al/Fe] result. Ba is discussed in
Section~\ref{heavy-abo}. For some elements (La, Mg, Nd, Y) no ratio could be
determined either because their lines appear too weak in the sun, or because the
solar spectrum range (5000--7000\AA) does not contain the lines we used in this
paper.

%%%%%%%%%%%%%%%%%%%%%%%%%%%%%%%%%%%%%%%%%%%%%%%%%%%%%% M67
\begin{table*}
\begin{minipage}[t]{\textwidth}
\caption{Literature abundance determinations for M~67 based on high resolution
spectroscopy.}             
\label{m67}      
\centering          
\renewcommand{\footnoterule}{}  
\begin{tabular}{l r r r r r r r}     
\hline\hline        
        & {\em Here} & 
        T00\footnote{\citet{taut00}, from 6 red clump stars.} & 
	S00\footnote{\citet{she00}, from 4 turn-off stars (we ignored blue
	stragglers).} & 
	Y05\footnote{\citet{yong05}, from 3 red clump stars.} & 
	R06\footnote{\citet{ran06}, from 8 dwarfs and 2 slightly evolved stars.} &
	P08\footnote{\citet{pace08}, from 6 main sequence stars.} & 
	S09\footnote{\citet{san09}, from 3 red clump giants and 6 dwarfs.} \\
\hline
R=$\lambda/\delta\lambda$ & 30000   & 30--60000 & 30000  & 28000   & 45000   & 100000	  & 50000    \\
S/N                       & 50--100 & $\geq$100 & 40--70 & 30--100 & 90--180 & $\simeq$80 & 100--300 \\
\hline
$[$Fe/H$]$  &  +0.05$\pm$0.02($\pm$0.10) & --0.03$\pm$0.04 & --0.05$\pm$0.01 &  +0.02$\pm$0.01 &  +0.03$\pm$0.01 &  +0.03$\pm$0.04 & 0.00$\pm$0.01 \\	   
$[$Al/Fe$]$ &  +0.03$\pm$0.02($\pm$0.05) &  +0.14$\pm$0.07 & ---	&  +0.17$\pm$0.01 & --0.06$\pm$0.04 & --0.03$\pm$0.11 & ---	      \\
$[$Ba/Fe$]$ &  +0.25$\pm$0.02($\pm$0.06)	    &  +0.07$\pm$0.12 &  +0.21$\pm$0.20 & --0.02$\pm$0.03 & --- 	    & ---	      & ---	      \\
$[$Ca/Fe$]$ & --0.16$\pm$0.03($\pm$0.03) &  +0.05$\pm$0.09 & --0.02$\pm$0.03 &  +0.07$\pm$0.02 &  +0.05$\pm$0.04 &  +0.03$\pm$0.07 & ---	   \\
$[$Co/Fe$]$ &  +0.08$\pm$0.06($\pm$0.06) &  +0.08$\pm$0.07 & ---	     & ---	       & ---		 & ---  	   & ---	   \\
$[$Cr/Fe$]$ &  +0.01$\pm$0.03($\pm$0.06) &  +0.10$\pm$0.09 & ---	     & ---	       & --0.01$\pm$0.04 &  +0.03$\pm$0.09 & ---	   \\
$[$La/Fe$]$ &  +0.05$\pm$0.06($\pm$0.09) &  +0.13$\pm$0.10 & ---	     &  +0.11$\pm$0.02 & ---		 & ---  	   & ---	   \\
$[$Mg/Fe$]$ &  +0.27$\pm$0.04($\pm$0.10) &  +0.10$\pm$0.06 & --0.11$\pm$0.08 &  +0.16$\pm$0.02 & --0.01$\pm$0.03 & ---  	   & ---	   \\
$[$Na/Fe$]$ &  +0.08$\pm$0.09($\pm$0.04) &  +0.19$\pm$0.08 & --0.05$\pm$0.05 &  +0.30$\pm$0.03 &  +0.04$\pm$0.07 & --0.02$\pm$0.07 & ---	   \\
$[$Ni/Fe$]$ &  +0.05$\pm$0.01($\pm$0.10) &  +0.04$\pm$0.08 &  +0.04$\pm$0.06 &  +0.08$\pm$0.02 & --0.01$\pm$0.04 & --0.02$\pm$0.07 & ---	   \\
$[$O/Fe$]$  &  +0.04$\pm$0.10($\pm$0.03) &  +0.02$\pm$0.06 & --0.02$\pm$0.08 &  +0.07$\pm$0.02 & --0.04$\pm$0.01 & --0.07$\pm$0.09 & ---	   \\
$[$Sc/Fe$]$ & --0.03$\pm$0.04($\pm$0.08) &  +0.10$\pm$0.06 & ---	     & ---	       & ---		 & ---  	   & ---	   \\
$[$Si/Fe$]$ &  +0.10$\pm$0.02($\pm$0.07) &  +0.10$\pm$0.05 & ---	&  +0.09$\pm$0.03 &  +0.03$\pm$0.04 & --0.03$\pm$0.06 & ---	      \\
$[$Ti/Fe$]$ & --0.04$\pm$0.06($\pm$0.06) &  +0.04$\pm$0.13 & ---	     &  +0.12$\pm$0.02 & --0.03$\pm$0.04 & --0.02$\pm$0.11 & ---	   \\
$[$V/Fe$]$  &  +0.15$\pm$0.13($\pm$0.14) &  +0.15$\pm$0.16 & ---	     & ---	       & ---		 & ---  	   & ---	   \\
$[$Y/Fe$]$  & --0.05$\pm$0.04($\pm$0.12) &  +0.01$\pm$0.14 & ---	     & ---	       & ---		 & ---  	   & ---	   \\
\hline             	     
\end{tabular}	   	     
\end{minipage}	   	     
\end{table*}	   
%%%%%%%%%%%%%%%%%%%%%%%%%%%%%%%%%%%%%%%%%%%%%%%%%%%%%% M67

%__________________________________________________________________

\section{Cluster-by-Cluster Discussion}
\label{sec-lit}

\subsection{Cr~110}

Collinder~110 is a poorly studied, intermediate-age OC located at
$\alpha_{\mathrm{J2000}}$$=$06:38:24 and $\delta_{\mathrm{J2000}}$$=$+02:01:00.
We could not find any high resolution spectroscopic study of this cluster in the
literature, but photometric studies have been conducted by \citet{tsa71},
\citet{daw98} and \citet{bra03}. Reddening, distance and ages determined by
these authors are included in Table~\ref{phot}. Concerning metallicity, while
the first two studies assume solar metallicity, \citet{bra03} find, as a result
of their synthetic diagram analysis, two equally good solutions, one at solar
metallicity and the other at slightly sub-solar metallicity (Z=0.008). A
re-evaluation of the same data by \citet{bra06} favours the slightly subsolar
value. 

Low resolution spectroscopy using the infrared calcium triplet by
\citet{ricardo} gave: [Fe/H]=--0.01$\pm$0.07~dex in the \citet{car97} scale,
[Fe/H]=0.0$\pm$0.3~dex in the \citet{zin84} scale and [Fe/H]=--0.19$\pm$0.21~dex
in the \citet{kra03} scale. Our determination of [Fe/H]=+0.03$\pm$0.02
($\pm$0.10)~dex is in good agreement with all these estimates, given the large
uncertainties involved in photometric and low-resolution spectroscopic
metallicity estimates. The other element ratios determined here have no
previous literature values to compare with, but the comparisons in
Section~\ref{sec-disc} show that they behave like expected for a solar
metallicity OC. Our radial velocity estimate for Cr~110,
$<$V$_r$$>$=41.0$\pm$3.8~km~s$^{-1}$ (Section~\ref{sec-vel}), is in good
agreement with the CaT value by \citet{ricardo} of 45$\pm$8~km~s$^{-1}$.

\subsection{NGC~2099 (M~37)}

NGC~2099 (M~37) is located in the Galactic anti-center direction in Auriga
$\alpha_{\mathrm{J2000}}$$=$05:52:18 and $\delta_{\mathrm{J2000}}$$=$+32:33:12.
Since it is near and it appears as a relatively rich and large cluster, it has
been photometrically studied by several authors to derive accurate magnitudes,
proper motions, and a census of variable stars \citep[for historical references
see][]{kalirai01}. All the papers that derived reddening, distance and age are
also cited in Table~\ref{phot}. Photometric studies generally attribute a solar
metallicity to NGC~2099 \citep[e.g.,][]{mer96}. Metallicity estimates based on
photometry can only be found in \citet{jan88}, who give [Fe/H]=0.09~dex;
\citet{mar05}, who give [M/H]$=$+0.05 $\pm$0.05 and \citet{kal04}, who give
Z$<$0.02.

Surprisingly, when considering the wealth of photometric studies, M~37 lacks any
specific low or high resolution study aimed at determining its chemical
composition. Our values therefore fill this gap, and show that in all
respects this cluster has a typical solar metallicity, with all element ratios
close to zero within the uncertainties. On the other hand, radial velocity
determinations for this cluster are quite abundant (Section~\ref{sec-vel}) and
appear in good agreement with our determination. 

\subsection{NGC~2420}
\label{lit-2420}

NGC~2420 ($\alpha_{\mathrm{J2000}}$$=$07:38:23 and
$\delta_{\mathrm{J2000}}$$=$+21:34:24) has always been considered the definitive
example of the older, moderately metal-deficient OC beyond the solar circle.
Several good quality imaging studies appeared already in the 60s and 70s
\citep[][to name a few]{sar62,west67b,can70,van70,mcc74,mcc78}, and more recent
photometries appeared in a variety of photometric systems \citep[the most cited
being][]{tw90}. Its intermediate status between the solar-metallicity OC near the
sun and the clearly metal-deficient population of globular clusters tagged it
early on as a potential transition object between the two populations, with
metallicity determinations --- both photometric and spectroscopic --- placing it
at an [Fe/H] value around the one of 47~Tuc
\citep[e.g.][]{pil80,coh80,can86,smi87}. More recent photometric works give
somewhat higher [Fe/H] values, ranging from --0.5 to --0.3~dex
\citep[e.g.][]{tw97,friel02,tw06}, but still significantly lower than the value of
[Fe/H]=--0.05$\pm$0.03 ($\pm$0.10)~dex we found here.

However, both \citet{coh80} and \citet{pil80} noted that NGC~2420 should be
significantly more metal rich than the Globular Clusters they analyzed, i.e.,
M~71 \citep{coh80} and 47~Tuc \citep{pil80}, by some $\simeq$0.5~dex. Since they
placed M~71 and 47~Tuc around [Fe/H]=--1.3, they consequently placed NGC~2420 at
[Fe/H]=--0.6. The resolution of their spectra (R$<$10000) was much lower than
ours, but if we trust their analysis in a relative sense, and consider more
recent metallicity estimates for 47~Tuc and M~71 \citep[--0.76 and --0.73
respectively,][]{har96}, we would then place NGC~2420 around [Fe/H]$\simeq$--0.1
or --0.2. Having said that, it is surprising that there are no modern high
resolution studies of a cluster that was considered so important in the past.
The highest spectral resolution employed to study NGC~2420 is R$\simeq$15000
\citep{smi87}, with a spectral coverage of only 200\AA, giving [Fe/H]=--0.57.
Only the preliminary work of \citet{fre02} has suggested a higher, slightly
subsolar [Fe/H] value for NGC~2420. We also note that our [Fe/H] brings NGC~2420
more in line with other OC in the Galactic trends discussed in
Section~\ref{sec-trend}. Also, we cannot ignore the similarity with the case of
NGC~7789 (Section~\ref{lit-7789}), where high resolution spectroscopy by 
\citet{taut05} and us provides a much higher abundance than the previous
photometric and low/medium-resolution studies. Clearly, further high resolution
spectroscopy with modern instruments, possibly with R$\simeq$50000 and
S/N$\simeq$100 is needed for this cluster.

\subsection{NGC~2682 (M~67)}
\label{lit-m67}

Among the old OC, M~67 ($\alpha_{\mathrm{J2000}}$$=$08:51:18,
$\delta_{\mathrm{J2000}}$$=$+11:48:00) is quite close to us, with low reddening
(Table~\ref{phot}) and solar metallicity, so it is one of the most studied open
clusters, and a good target to look for solar twins and analogs \citep{pas08}.
Since the first pioneering studies at the beginning of XX century, we have a few
hundred papers published to date \citep[see][for more
references]{bur86,car96,yad08}. Therefore, we have included M~67 in our sample
because it acts as a fundamental comparison object, that enables us to place our
measurements in a more general framework. 

Among the vast literature on M~67, there are several determinations of its
metallicity, with various methods \citep[e.g.][to name a
few]{dem80,coh80,foy81,jan84,bur84,bur86,bro87,gar88,cay90,hob91,fri92,friel93,jan94,friel02,bal07,mar05},
all typically converging to a solar value. high resolution abundance
determinations have been derived for both giants and dwarfs, with many studies
devoted to light elements such as lithium and beryllium and their implications
for mixing theories \citep{pas97,jon99,ran07}. 

Table~\ref{m67} shows a comparison of our results with some of the most recent
high resolution (R$\geq$20000) determinations
\citep{taut00,she00,ran06,pace08,san09}. The overall comparison is extremely
satisfactory for all elements, except maybe for Mg, Na, Ba and Ca (see also
Section~\ref{ato-alpha}). For Mg, Na and Ba the large spread in literature
demonstrates the difficulties in measuring these elements. For Ca we see that our
value is marginally lower than other literature deterrminations. As explained in
Section~\ref{ato-alpha}, this is most probably due to the large uncertainties on
the Calcium log~$gf$ values. 

%%%%%%%%%%%%%%%%%%%%%%%%%%%%%%%%%%%%%%%%%%%%%%%%%%%%%% N7789
\begin{table}
\begin{minipage}[t]{\columnwidth}
\caption{Literature abundance determinations for NGC~7789 based on
high resolution spectroscopy.}             
\label{7789}      
\centering          
\renewcommand{\footnoterule}{}  
\begin{tabular}{l r r}     
\hline\hline        
        & {\em Here} & 
        T05\footnote{\citet{taut05}, from 6 giants and 3 red clump giants.} \\
\hline
R=$\lambda/\delta\lambda$ & 30000   & 30000 \\
S/N                       & 50--100 & $\geq50$ \\
\hline
$[$Fe/H$]$  &  +0.04$\pm$0.07($\pm$0.10) & --0.04$\pm$0.05 \\	   
$[$Al/Fe$]$ & --0.03$\pm$0.09($\pm$0.05) &  +0.18$\pm$0.08 \\
$[$Ca/Fe$]$ & --0.18$\pm$0.09($\pm$0.03) &  +0.14$\pm$0.07 \\
$[$Co/Fe$]$ & --0.02$\pm$0.01($\pm$0.06) &  +0.09$\pm$0.14 \\
$[$Cr/Fe$]$ &  +0.01$\pm$0.05($\pm$0.06) & --0.05$\pm$0.09 \\
$[$Mg/Fe$]$ &  +0.22$\pm$0.07($\pm$0.10) &  +0.18$\pm$0.07 \\
$[$Na/Fe$]$ & --0.05$\pm$0.13($\pm$0.04) &  +0.28$\pm$0.07 \\
$[$Ni/Fe$]$ & --0.01$\pm$0.01($\pm$0.10) & --0.02$\pm$0.05 \\
$[$O/Fe$]$  &  +0.16$\pm$0.06($\pm$0.03) & --0.07$\pm$0.09 \\
$[$Sc/Fe$]$ &  +0.08$\pm$0.02($\pm$0.08) & --0.02$\pm$0.07 \\
$[$Si/Fe$]$ & --0.01$\pm$0.02($\pm$0.07) &  +0.14$\pm$0.05 \\
$[$Ti/Fe$]$ & --0.03$\pm$0.09($\pm$0.06) & --0.03$\pm$0.07 \\
$[$V/Fe$]$  & --0.01$\pm$0.09($\pm$0.14) &  +0.09$\pm$0.12 \\
$[$Y/Fe$]$  &  +0.08$\pm$0.09($\pm$0.12) &  +0.13$\pm$0.13 \\
\hline             	     
\end{tabular}	   	     
\end{minipage}	   
\end{table}	   
%%%%%%%%%%%%%%%%%%%%%%%%%%%%%%%%%%%%%%%%%%%%%%%%%%%%%% N7789

\subsection{NGC~7789}
\label{lit-7789}

NGC7789 ($\alpha_{\mathrm{J2000}}$$=$23:57:24 and
$\delta_{\mathrm{J2000}}$$=$+56:42:30, or $l$$=$115.53 and $b$$=$--5.39) is a
rich and intermediate-age OC, with a well defined giant branch,
a well-populated main-sequence turnoff, and a substantial population  of blue
stragglers \citep{mcn80,twa85,mil94}. Several photometric studies have been
carried out \citep[some examples
are][]{kun23,red54,bur58,jan77,mar94,jah95,gim98b,val00,bar04,bra05}
and its parameters are reasonably well known.

%%%%%%%%%%%%%%%%%%%%%%%%%%%%%%%%%%%%%%%%%%%%%%%%%%%%%% Literature HiRes
\begin{table*}
\begin{minipage}[t]{\textwidth}
\caption{Literature sources and [Fe/H] values for high resolution (R$\geq$15000)
based abundance ratios of old OC.}             
\label{tab-hires}      
\centering          
\renewcommand{\footnoterule}{}  
\begin{tabular}{l r l l r l}     
\hline\hline
Cluster & [Fe/H] & Reference & Cluster & [Fe/H] & Reference \\        
\hline
Be 17    & --0.10 & \citet{fri05}   & NGC 2141 & --0.26 & \citet{yong05}  \\
Be 20    & --0.61 & \citet{yong05}  &          &   0.00 & \citet{jac09}   \\
         & --0.30 & \citet{sestito} & NGC 2158 & --0.03 & \citet{jac09}   \\ 
Be 22    & --0.32 & \citet{vil05}   & NGC 2243 & --0.48 & \citet{gra94}   \\	     
Be 25    & --0.20 & \citet{car07}   & NGC 2324 & --0.17 & \citet{bra08}   \\	     
Be 29    & --0.44 & \citet{carr04}  & NGC 2360 &  +0.07 & \citet{ham00}   \\
         & --0.18 & \citet{yong05}  &          &  +0.04 & \citet{smi08}   \\
         & --0.31 & \citet{sestito} & NGC 2420 & --0.57 & \citet{smi87}   \\
Be 31    & --0.40 & \citet{yong05}  & NGC 2447 &  +0.03 & \citet{ham00}   \\
Be 32    & --0.29 & \citet{bra08}   & NGC 2477 &  +0.07 & \citet{bra08}   \\
Be 66    & --0.48 & \citet{vil05}   &          & --0.01 & \citet{smi08}   \\
Be73     & --0.22 & \citet{car07}   & NGC 2506 & --0.20 & \citet{car04}   \\
Be75     & --0.22 & \citet{car07}   & NGC 2660 &  +0.04 & \citet{bra08}   \\
Blanco 1 &  +0.04 & \citet{for05}   & NGC 3532 &  +0.04 & \citet{smi08}   \\
Cr 261   & --0.22 & \citet{fri03}   & NGC 3680 & --0.04 & \citet{pace08}  \\
         & --0.03 & \citet{car05}   &          &  +0.04 & \citet{smi08}   \\
         & --0.03 & \citet{des07}   & NGC 3960 &  +0.02 & \citet{bra08}   \\
         &  +0.13 & \citet{sestito} & NGC 5822 &  +0.04 & \citet{smi08}   \\ 
Hyades   &  +0.13 & \citet{ses03}   & NGC 6134 &  +0.15 & \citet{car04}   \\ 
         &  +0.13 & \citet{pau03}   &          &  +0.12 & \citet{smi08}   \\
         &  +0.13 & \citet{des06}   & NGC 6253 &  +0.46 & \citet{carr07}  \\
IC 2391  & --0.03 & \citet{ran01}   &          &  +0.36 & \citet{ses07}   \\
IC 2602  & --0.05 & \citet{ran01}   & NGC 6281 &  +0.05 & \citet{smi08}   \\
IC 2714  &  +0.12 & \citet{smi08}   & NGC 6475 &  +0.14 & \citet{ses03}   \\
IC 4756  & --0.15 & \citet{jac07}   & NGC 6633 &  +0.07 & \citet{smi08}   \\
         &  +0.04 & \citet{smi08}   & NGC 6791 &  +0.40 & \citet{pet98}   \\
IC 4651  &  +0.11 & \citet{car04}   &          &  +0.35 & \citet{ori06}   \\
         &  +0.10 & \citet{pas04}   &          &  +0.39 & \citet{carr06}  \\
         &  +0.12 & \citet{pace08}  &          &  +0.47 & \citet{carr07}  \\
M 11     &  +0.10 & \citet{gon00}   &          &  +0.30 & \citet{boe09}   \\
M 34     &  +0.07 & \citet{sch03}   & NGC 6819 &  +0.09 & \citet{bra01}   \\
M 67     & --0.03 & \citet{taut00}  & NGC 6939 &   0.00 & \citet{jac07}   \\
         & --0.01 & \citet{yong05}  & NGC 7142 &  +0.08 & \citet{jac08}   \\
         &  +0.03 & \citet{ran06}   & NGC 7789 & --0.04 & \citet{taut05}  \\
         &  +0.03 & \citet{pace08}  & Pleiades & --0.03 & \citet{ran01}   \\
Mel 66   & --0.38 & \citet{gra94}   &	       &  +0.06 & \citet{geb08}   \\
         & --0.33 & \citet{sestito} & Praesepe &  +0.04 & \citet{fri92}   \\
Mel 71   & --0.30 & \citet{bro96}   &	       &  +0.12 & \citet{pace08}  \\
NGC 188  &  +0.01 & \citet{ran03}   & Rup 4    & --0.09 & \citet{car07}   \\
NGC 1817 & --0.07 & \citet{jac09}   & Rup 7    & --0.26 & \citet{car07}   \\
NGC 1883 & --0.20 & \citet{vil07}   & Saurer 1 & --0.38 & \citet{carr04}  \\
         & --0.01 & \citet{jac09}   & Tom 2    & --0.45 & \citet{bro96}   \\
NGC 2112 & --0.10 & \citet{bro96}   &	       & --0.28 & \citet{frin08}  \\
\hline\hline             	  
\end{tabular}	   
\end{minipage}	   
\end{table*}	   
%%%%%%%%%%%%%%%%%%%%%%%%%%%%%%%%%%%%%%%%%%%%%%%%%%%%%% Literature HiRes

Abundance determinations through photometry and low/medium-resolution
spectroscopy all give sub-solar values around [Fe/H]$\simeq$--0.2
\citep{pil85,friel93,sch01,friel02}, i.e., much lower than our
[Fe/H]=0.04$\pm$0.07 ($\pm$0.10)~dex. However, a more conforting comparison with
\citet{taut05} is shown in Table~\ref{7789}. Their spectra have resolution and S/N
similar to ours, and most abundance ratios in common show an excellent agreement.
Minor discrepancies arise for some elements such as Ca (but see the discussions in
Sections~\ref{ato-alpha} and \ref{lit-m67}), Al (but they used only one doublet
while we used four), Na and O. Since they do not list their log~$gf$ values, and
other ingredients of the abundance analysis were similar to ours, we cannot
explain the Na-O discrepancies, but we suspect that they must be due log~$gf$
differences. 

%__________________________________________________________________

\section{Abundance Ratios Discussion}
\label{sec-disc}

We compare our abundance ratios with data from the literature, assembled as
follows. For the Milky Way field stars, we use the Thick and Thin Disc
measurements from \citet{red03} and \citet{red06}, who performed homeogeneous
abundance calculations of a few hundred F/G dwarfs selected from the Hipparcos
catalogue. We added abundance ratios, based on high resolution spectroscopy, for
57 old OC from various literature sources (Table~\ref{tab-hires}). When more
than one determination was available for one cluster, we simply plotted them all
to give a realistic idea of the uncertainties involved in the compilation. 

%%%%%%%%%%%%%%%%%%%%%%%%%%%%%%%%%%%%%%%%%%%%%%%%% Ratio Iron-peak
\begin{figure}
\centering
\includegraphics[width=\columnwidth]{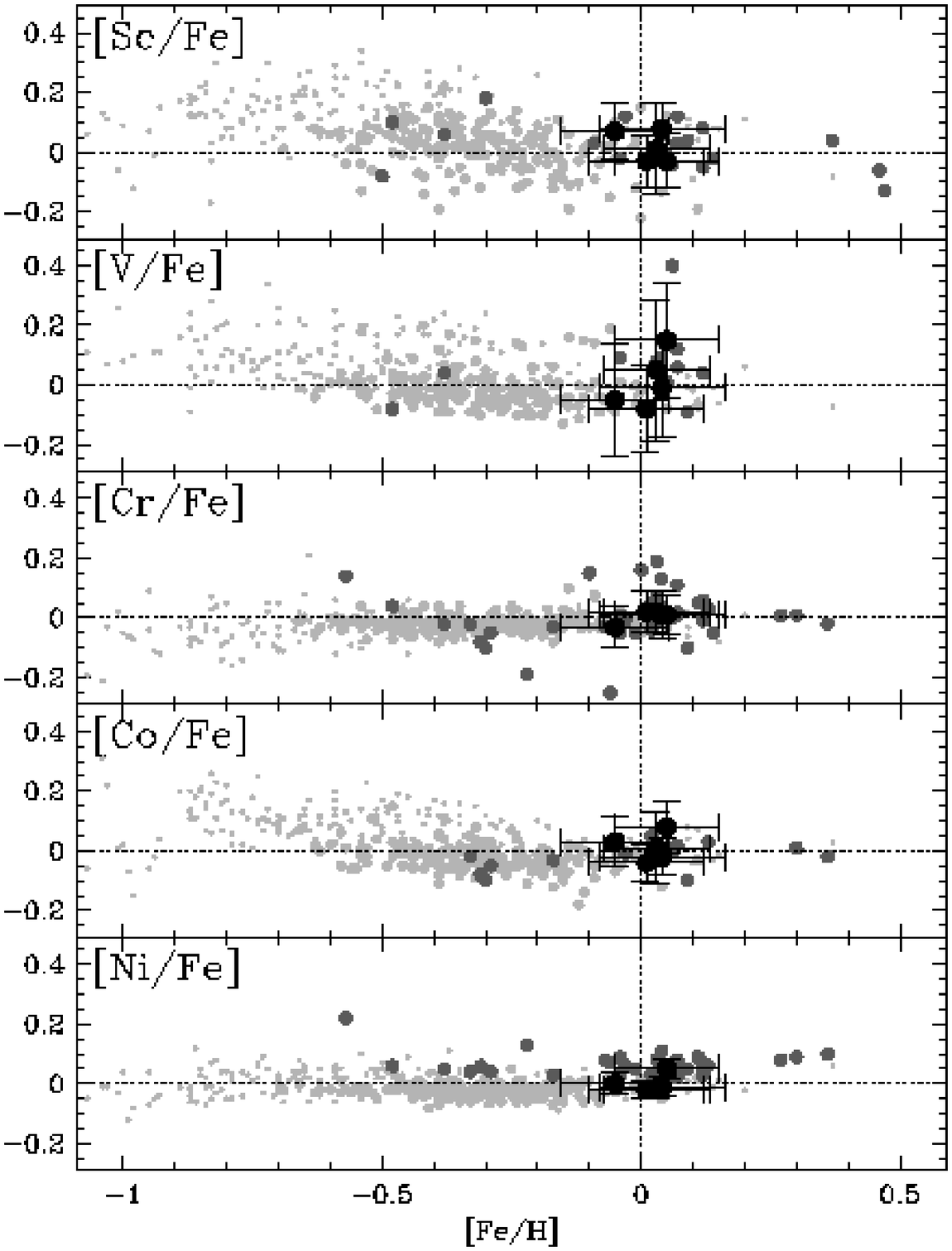}

\caption{Comparison between our iron-peak element results (large black dots),
the high resolution measurements of other OC listed in Table~\ref{tab-hires}
(large dark grey dots) and field stars belonging to the Thin Disc \citep[light
grey dots,][]{red03} and to the Thick Disc \citep[tiny light grey
dots,][]{red06}. Errorbars on our results are the quadratic sum of internal
uncertainties and uncertainties due to the choice of stellar parameters
(Section~\ref{sec-abo}).}

\label{ra_iron}
\end{figure}
%%%%%%%%%%%%%%%%%%%%%%%%%%%%%%%%%%%%%%%%%%%%%%%%% Ratio Iron-peak

\subsection{Iron-peak Elements Ratios}
\label{iron-abo}

When compared with literature (Figure~\ref{ra_iron}), our iron-peak elements
appear all solar and in good agreement with the results for the Disc and other
OC. In particular, cobalt and chromium have the best agreement and smallest
spreads. Although Sc, V and Co are known to possess HFS that may lead to an
increased scatter and an overestimated ratio, they do not appear significantly
different from solar for our target stars, so we did not attempt any detailed
HFS analysis. Nevertheless, the effect of increased scatter and overestimated
abundance are visible, expecially for vanadium, both in our data and in the
Discs stars, as well as in the other clusters from the literature. 

A puzzling effect is seen in Figure~\ref{ra_iron} in the [Ni/Fe] ratio. All the
data from Disc stars are very close to solar ($<$[Ni/Fe]$>$=--0.02$\pm$0.02), and
so are our determinations ($<$[Ni/Fe]$>$=0.00$\pm$0.03), but the other OC high
resolution data appear slightly enhanced ($<$[Ni/Fe]$>$=0.06$\pm$0.04), lying
systematically above the Disc ones. Such a $\simeq$0.05~dex offset is well within
the uncertainties of abundance determinations in general, but since it appears
systematic in nature, we are still left without a clear explanation. Our [Ni/Fe]
ratios anyway are a bit lower than the other OC determinations, although still
compatible within the uncertainties.

%%%%%%%%%%%%%%%%%%%%%%%%%%%%%%%%%%%%%%%%%%%%%%%%% Ratio Iron-peak
\begin{figure}
\centering
\includegraphics[width=\columnwidth]{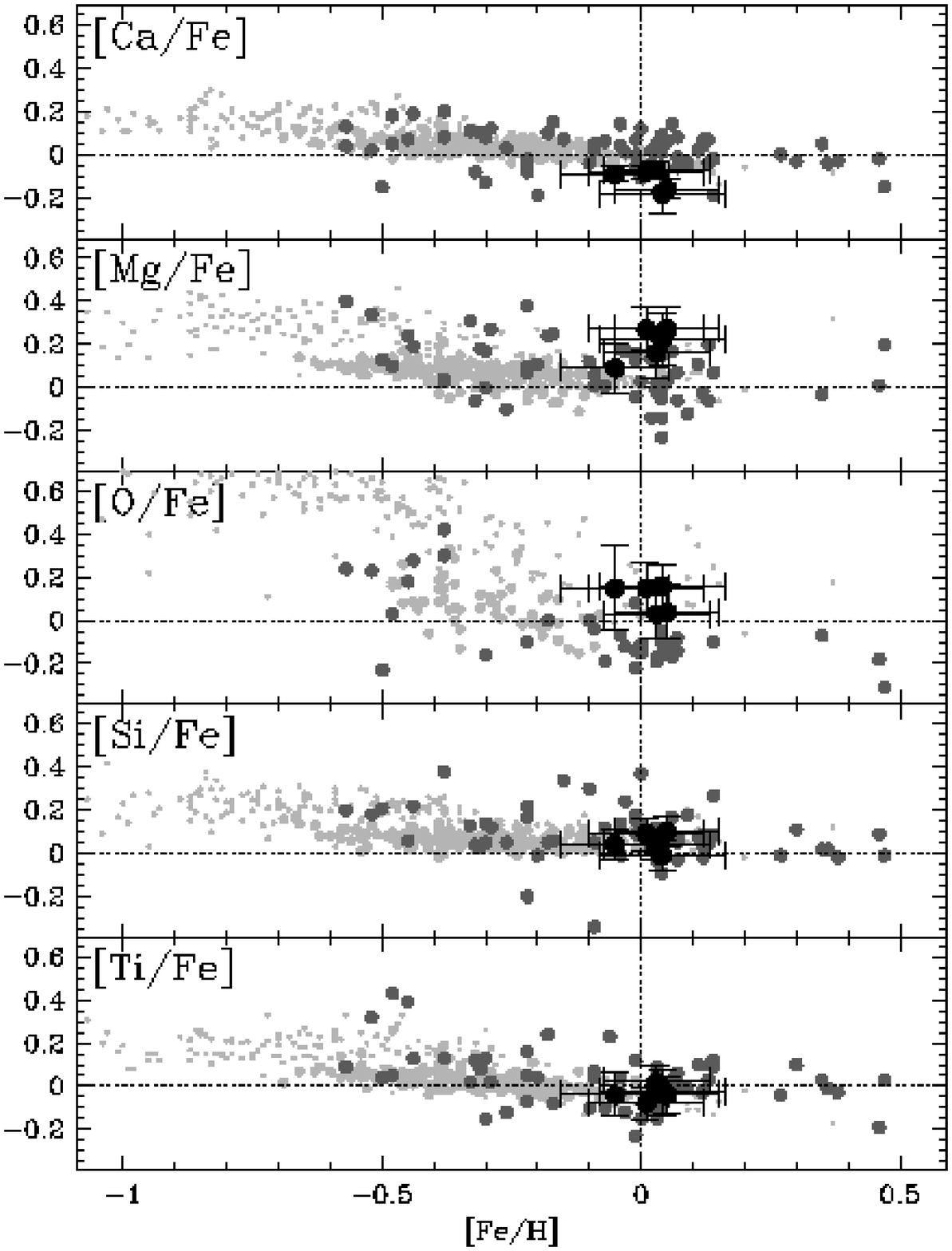}

\caption{Comparison between our $\alpha$-elements ratios and the literature
ones. Symbols are the same as in Figure~\ref{ra_iron}.}

\label{ra_alfa}
\end{figure}
%%%%%%%%%%%%%%%%%%%%%%%%%%%%%%%%%%%%%%%%%%%%%%%%% Ratio Iron-peak

\subsection{$\alpha$-elements Ratios}
\label{alfa-abo}

We obtained abundances of Ca, Mg, O, Si and Ti. As can be seen from
Figure~\ref{ra_alfa}, Si and Ti appear practically solar, within the respective
uncertainties, and in very good agreement with literature determinations for
both the Discs stars and the other OC. O, Ca and Mg give instead marginally
discrepant enhancements. 

For O, we note that the spread both in our data and in the literature is greater
that in any other $\alpha$-element. This is partly due to the well known
problems of determining O from the 6300\AA\  lone line, or the 6363\AA\  weak line,
or from the IR triplet at 7770\AA, that requires NLTE corrections. Moreover, some
old literature work uses Solar reference abundances reaching as high as 8.93, which
can explain some of the lowest [O/Fe] literature estimates. Given the large
spread, the tendency of our [O/Fe] measurements to lie on the upper envelope of the
other OC high resolution data is probably irrelevant. Solar O enhancements would
probably be more in line with the other $\alpha$-elements, while the sub-solar
values found generally in literature point towards Wolf-Rayet as additional
contributors of O, with a stronger metallicity dependence of the O yields
\citep{mcw08}.

In the case of Ca, our values are marginally inconsistent with the bulk of
field and OC literature determinations. A few literature measurements of OC ratios
are however as low as our values. These inconsistencies could be explained with
the large uncertainties in the literature log$gf$ values for Calcium lines
($\sim$0.2~dex, see discussion in Section~\ref{ato-alpha}). Given these large
additional uncertainties, we finally concluded that [Ca/Fe] is basically
compatible with solar values in all the clusters examined.

Concerning Mg, we know already that the log~$gf$ values of some lines are still not
very well determined (Section~\ref{ato-alpha}). We also know \citep{gra99} that
some lines require NLTE corrections. We could find no correction factors for the
lines we were able to measure in our spectra, but we noticed that those lines
examined by \citet{gra99} which have $\chi_{ex}$ similar to our lines, require
NLTE corrections of about +0.1--0.5~dex. This correction would make our [Mg/Fe]
values even higher, reaching an enhancement of 0.2--0.6~dex with respect to solar.
Another possibility is that our lines have a non negligible HFS, because they are
dominated by odd isotopes, but we could find no further information in the
literature. We could only notice that other authors find such relatively high
values of [Mg/Fe] in OC \citep[such as][]{bra08}.

When the average $[$$\alpha$/Fe$]$ values are calculated, however, all the
programme stars and the cluster averages appear perfectly compatible with solar,
within relatively small uncertainties, as expected (see Tables~\ref{abotab} and
\ref{tab-cluster}). The [$\alpha$/Fe] ratio is also discussed further in
Section~\ref{sec-trend}.

%%%%%%%%%%%%%%%%%%%%%%%%%%%%%%%%%%%%%%%%%%%%%%%%% Ratio s-process
\begin{figure}
\centering
\includegraphics[bb=20 20 600 650,width=\columnwidth]{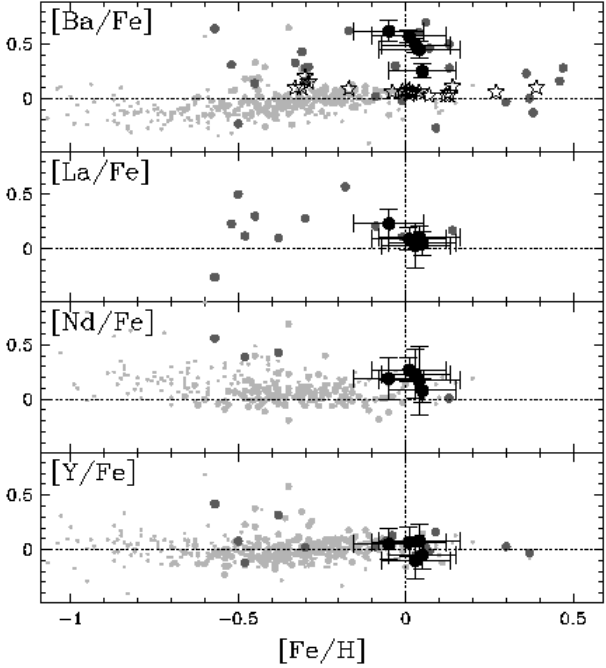}

\caption{Comparison between our s-process elements ratios and the literature
ones. Symbols are the same as in Figure~\ref{ra_iron}, except for the black
star-like symbols in the top [Ba/Fe] panel, which represent the revision of Ba
abundances with spectral synthesis performed by \citet{dor09}.}

\label{ra_esse}
\end{figure}
%%%%%%%%%%%%%%%%%%%%%%%%%%%%%%%%%%%%%%%%%%%%%%%%% Ratio s-process

\subsection{Heavy Elements Ratios}
\label{heavy-abo}

We measured the heavy s-process elements Ba, La and Nd and the light s-process
element Y. La does not require any synthesis to take into account HFS, since the
three lines we used are always in the linear part of the curve of growth, and in
fact Figure~\ref{ra_esse} shows good agreement with the sparse literature
values. Yittrium and neodimium are also in agreement with the literature data,
although the measurements of Nd in OC are scarce and scattered. Our values of
[Nd/Fe] have a tendency of being towards the upper envelope of the Dics stars,
but this is not significant if we consider the large uncertainties involved (see
also the discussion in Section~\ref{ato-heavy}). 

Concerning Ba, we find high values both in our programme stars and in the sun
itself (Section~\ref{sec-sun}). The same result has been found by other authors
\citep[e.g.,][]{bra08}. While a detailed study of the barium abundance is out of
the scope of the present paper, we note that very recently \citet{dor09} used a
detailed HFS analysis of barium in OC and showed that the
overabundance can be thus reduced roughly by $\sim$0.2~dex. Looking at
Figure~\ref{ra_esse}, we see that in fact our [Ba/Fe] are on the high side of
the OC data, which in turn have a huge spread. Data from \citet{dor09}, who
revised the Ba abundances for 20 OC using spectral synthesis to take HFS into
account, are towards the lower envelope of the OC abundances (open stars in
Figure~\ref{ra_esse}). As can be seen, some [Ba/Fe] ehnancement remains in their
high quality data, that is apparently well correlated with the cluster ages (see
their Figure~1). Still no clear explanation is available, since the current
evolution models and yields do not reproduce the data correctly at young ages,
where the enhancement is higher and more uncertain (up to [Ba/Fe]$\simeq$0.6~dex
for ages around 10$^8$). In summary, most of the [Ba/Fe] enhancement we see in
our measurements should be due to HFS effects, but some part of it could be real
\citep[up to 0.2~dex, see Fig.~2 by][]{dor09}. A hint of a descending slope of
[Ba/Fe] in OC appears, that is not appaerent among Disc stars. Further studies
such as \citet{dor09} are necessary on large samples of Disc and OC stars to get
firmer constraint on the chemical evolution of Ba in the Galactic Dics.

%%%%%%%%%%%%%%%%%%%%%%%%%%%%%%%%%%%%%%%%%%%%%%%%% Ratio Na-Al
\begin{figure}
\centering
\includegraphics[bb=20 20 600 380,clip,width=\columnwidth]{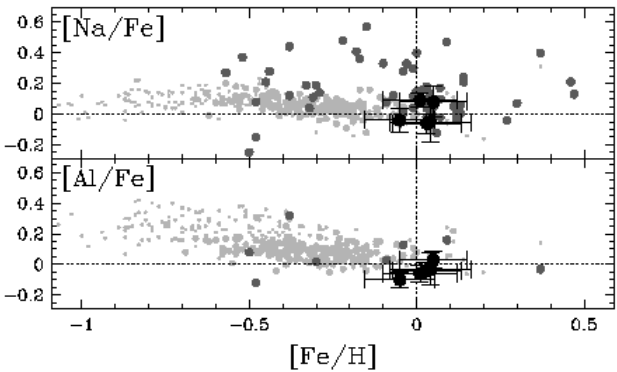}

\caption{Comparison between our [Na/Fe] and [Al/Fe] ratios and the literature
ones. Symbols are the same as in Figure~\ref{ra_iron}.}

\label{ra_naal}
\end{figure}
%%%%%%%%%%%%%%%%%%%%%%%%%%%%%%%%%%%%%%%%%%%%%%%%% Ratio Na-Al

\subsection{Ratios of Na and Al and Anticorrelations}

We also derived Na and Al abundances, since these elements are quite easy to
measure in OC and there is a vast body of literature measurements to compare with.
Figure~\ref{ra_naal} shows that for Al there is a general agreement between our
data, the Disc ratios and the OC ratios, although there is a large scatter in the
literature data. NLTE corrections for the four Al doublets and the type of stars
studied here could not be found in the literature. \citet{bau97} give
corrections for hotter (T$_{\rm{eff}}$$>$5000) and higher gravity ($\log g$$>$3.5)
stars, that suggest that either the corrections are negligible, or they are
slightly negative at lower temperatures and gravities.

In the case of Na, the spread in the literature data is even larger, and there
are both data points with significant Na enhancement and points with typical
solar values. Part of the scatter depends on the need for NLTE corrections.
According to \citet{gra99}, the 5682--5688\AA\  and the 6154--6160\AA\  doublets
at EW$\simeq$100~m\AA\  both require corrections of about $\leq$0.05--0.10~dex,
for solar stars like the ones considered here. The NLTE corrected abundances
should be higher than the LTE uncorrected ones: this should make our [Na/Fe] LTE
measurements in better agreement with literature measurements. If the observed
enhancement in [Na/Fe] should prove to be real for OC, this would set OC stars
completely apart form Disc stars \citep{des09}. If the large spread will also
prove to be intrinsic, this would suggest the possibility of light elements
chemical anomalies similar, although much less pronounced, to the ones found in
Globular Clusters. 

%%%%%%%%%%%%%%%%%%%%%%%%%%%%%%%%%%%%%%%%%%%%%%%%% Anticorrelations
\begin{figure}
\centering
\includegraphics[width=\columnwidth]{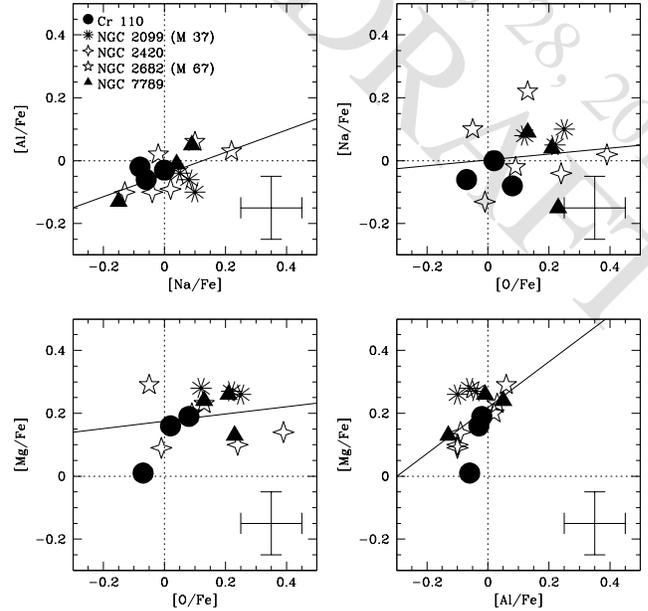}

\caption{A search for (anti)-correlations of Al, Mg, Na and O among our targets
stars. The four panels show different planes of abundance ratios, where stars
belonging to each cluster are marked with different symbols. Dotted lines show
solar values, solid lines show linear regressions and the typical uncertainty
($\sim$0.1~dex) is marked on the lower right corner of each panel.}

\label{anticorr}
\end{figure}
%%%%%%%%%%%%%%%%%%%%%%%%%%%%%%%%%%%%%%%%%%%%%%%%% Anticorrelations

In fact Al and Na, together with Mg and O (and C and N) show puzzling
(anti)-correlations in almost all of the studied Galactic Globular Clusters
\citep{gra04}. The most interesting fact concerning the chemical anomalies, is
that they have never been found outside globular clusters. They are not present
in the field populations of the Milky Way and its surrounding dwarf galaxies,
and they only recently have been found in Fornax and LMC clusters
\citep{let06,joh06,muc09}. Recent dedicated searches for anti-correlations in OC
are the one by \citet{mar09}, on the strength of CH and CN bands in NGC~188,
NGC~2158 and NGC~7789, the one by \citet{smi08}, based on high resolution
spectroscopy of C, N, O, Na and $^{12}$C/$^{13}$C and the one by \citet{des09},
that compiles and homogenizes literature Na-O high resolution data. No clear-cut
sign of anti-correlation has been found yet. 

If we build the usual (anti)-correlation plots for our five OC
(Figure~\ref{anticorr}), we find no clear sign of chemical anomalies, and in all
four plots the spread of each ratio is still compatible with the typical
uncertainty of our measurements, which is of the order of 0.1--0.2~dex,
depending on the element. In particular, in the [O/Fe]--[Na/Fe] plane, all 15
stars roughly occupy the solar region around zero that in Figure~5 by
\citet{car06} contains normal stars only \citep[see also][]{des09}. A possible
exception to this total absence of correlations is the [Na/Fe]--[Al/Fe] plane,
where a hint of a correlation can be noticed. Statistically, this is not
significant and small variations in T$_{\rm{eff}}$ could induce a similar weak
correlation. For this reason, anti-correlations are usually a more robust sign
of chemical anomalies. Still, in light of the discussion by \citet{smi08} about
the Na-O anti-correlation, our Na-Al results is suggestive. If further studies
will show that some chemical anomalies of these elements are present in Galactic
OC, our analysis \citep[together with that of][]{smi08} shows that
they must be of a much smaller extent than in Globular Clusters: 0.2--0.3~dex at
most in the [Al/Fe], [Mg/Fe], [Na/Fe] and [O/Fe] for the kind of clusters
studied here. In any case, the lack of relations in OC would point towards one
or more of the following environmental causes for the presence of
anticorrelations in globular clusters: {\em (i)} relatively low metallicity
(below solar); {\em (ii)} dense environment; {\em (iii)} total cluster mass of
the order of $\sim$10$^{4}$M$_{\odot}$ or more; {\em (iv)} undisturbed
environment \citep[e.g. away from the Disc, see also][]{eu06}.

\section{Galactic Trends}
\label{sec-trend}

\begin{figure}
\centering
\includegraphics[width=\columnwidth]{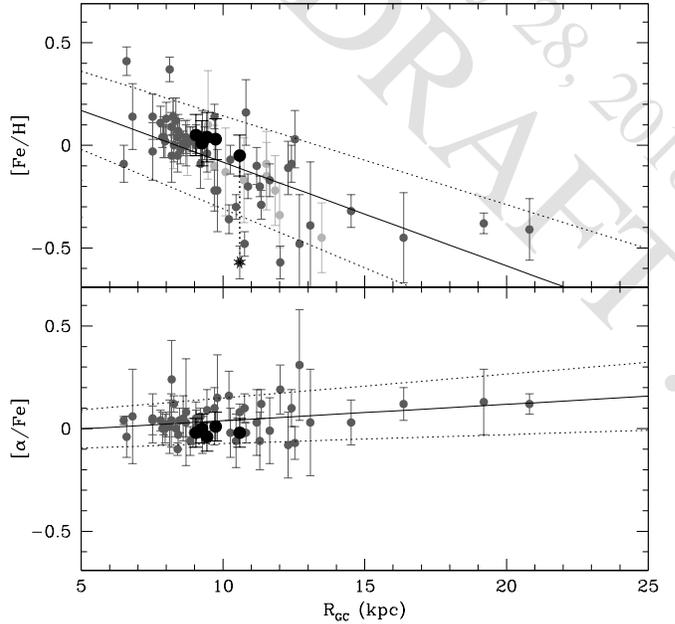}

\caption{Trends of [Fe/H] (top panel) and [$\alpha$/Fe] (bottom panel) with
Galactocentric Radius. Light grey dots in the top panel are OC from
\citet{friel02}, grey dots are the OC compiled in Table~\ref{tab-hires} and
black dots are our data. NGC~2420 estimates by us and \citet{smi87} are
connected with a thick dotted line. A linear fits with uncertainty is drawn
across all points in both panels.}

\label{tr_rgc}
\end{figure}

\begin{figure}
\centering
\includegraphics[width=\columnwidth]{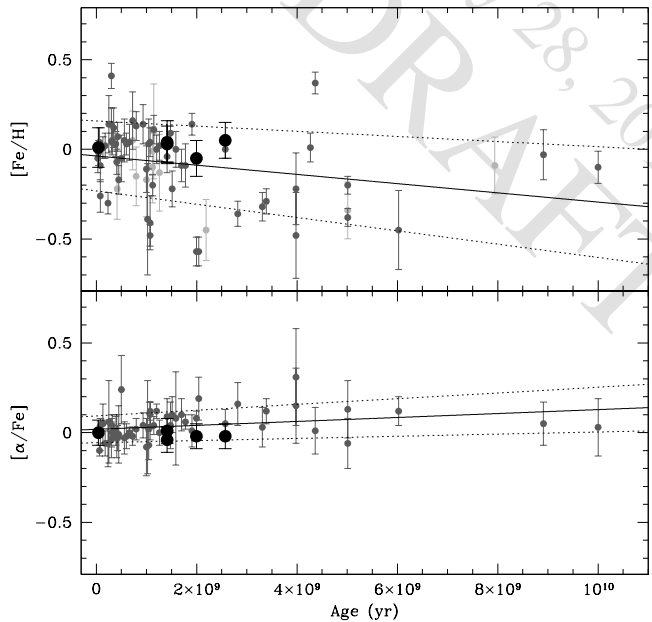}

\caption{Trends of [Fe/H] (top panel) and [$\alpha$/Fe] (bottom panel) with
cluster ages. Symbols are the same as in Figure~\ref{tr_rgc}.  }

\label{tr_age}
\end{figure}

As said earlier, OC are the fundamental test particles for the study of the
chemical evolution of the Galactic Disc, and as such, they produce two of the
strongest constraints on chemical models: the Galactic radial trends and the
Age-Metallicity Relation (AMR). Since a careful homogeneization of literature
data (including not only element ratios, but also ages and Galactocentric radii)
is out of the scope of the present paper, we have used the literature data of
Table~\ref{tab-hires}, averaging all estimates for a single cluster together. We
complemented with data by \citet{friel02} for those OC lacking high resolution
measurements. Using 28 OC in common between the two datasets, we found an
average offset of [Fe/H]=--0.16$\pm$0.13~dex, in the sense that the measurements
by \citet{friel02} are on average smaller than the ones based on high
resolution. We corrected \citet{friel02} data by this amount before plotting
them. We extracted OC ages from the compilation of \citet{dia02}\footnote{We are
aware that at least in the case of NGC~6791, the age given by \citet{dia02} is
quite different from other literature estimates (citations in
Tabls~\ref{tab-hires}), being lower by at least 2~Gyr. However, building a
homogeneous age scale is a non-trivial task, and it is beyond the scope of the
present paper.}, and for the Galactocentric Radii (R$_{\rm{GC}}$) we used
primarily \citet{friel02}, complemented by the WEBDA, and filled in the few
missing clusters with data from the papers of Table~\ref{tab-hires}. The
resulting radial trends and AMR for [Fe/H] and [$\alpha$/Fe] (computed as in
Section~\ref{abo-calc}) are plotted in Figures~\ref{tr_rgc} and \ref{tr_age} and
discussed below. 

%%%%%%%%%%%%%%%%%%%%%%%%%%%%%%%%%%%%%%%%%%%%%%%%% Radial Trends
%%%%%%%%%%%%%%%%%%%%%%%%%%%%%%%%%%%%%%%%%%%%%%%%% Radial Trends

\subsection{Trends with Galactocentric Radius}

The trend of abundances with Galactocentric radius R$_{\rm{GC}}$ gives a  strong
constraint for the models of Galactic chemical evolution as far as the Disc
formation mechanism is concerned\footnote{A far stronger constraint would be the
variation of this trend with age. Such a study is at the moment not possible,
given the small number of clusters studied with high resolution in a homogenous
way.}. It is now widely accepted \citep[see for
example][]{tw97,friel02,yong05,sestito,mag09} that there is a clear trend of
decreasing metallicity, measured as [Fe/H], with increasing R$_{\rm{GC}}$. Such
a trend is clearly detected not only in OC, but also in field stars (B-stars and
Cepheids), H~II regions and planetary nebulae\footnote{Although we know of at
leas one dataset in which PNe show flat trends of oxygen and neon abundances
with R$_{\rm{GC}}$ \citep{sta06}.} \citep[see][for some review of literature
data]{chi01,and04,yong05,lem08}. 

The first large studies of homogeneous OC abundances \citep[summarized in the
review by][]{fri95}, found that old OC (older than the Hyades) were extending
outwards in the Disc, much farther out than young OC, and they found a well
defined slope out to R$_{\rm{GC}}$$\simeq$16~kpc. The spread around this slope
was at the time around $\sim$0.2~dex, i.e., generally compatible with (or
perhaps slightly larger than) the measurement uncertainties. Such a slope comes
naturally in most Galactic chemical evolution models 
\citep{tos82,tos88,mat89,tos96,chi01,and04,col09,mag09}, when different star
formation and infall rates are assumed for the inner and outer Disk. To
reproduce most of the observational constraints, a differential Disc formation
mechanism is often assumed, either with the inner Disk formed first and then
growing in radius (inside-out formation) or with the whole Disk evolving
simultaneously, but with a much more intense (and sometimes fast) evolution in
the central, denser parts. A prediction of all models is that the metallicity
gradient should change with time \citep[although different models predict very
different time changes,][]{tos96} and some predict that it should flatten out at
large radii. Indeed, the first studies of anticenter and distant clusters
\citep{carr04,yong05,car07,sestito} showed that after
R$_{\rm{GC}}$$\simeq$12~kpc, the relation flattens around a value of
[Fe/H]$\simeq$--0.3~dex. 

Not much can be said with the present data about the slope variation with time,
but the exact value of the slope has been matter of some debate. As said,
earlier studies found a value around --0.09$\pm$0.01~dex~kpc$^{-1}$, or
--0.07$\pm$0.01~dex~kpc$^{-1}$ with the strictly homogenous measurements by
\citet{friel02}. An alternative interpretation of a different data compilation
\citep{tw97}, describes the trend as two disjoint plateaux, one around solar
metallicity and comprising OC within R$_{\rm{GC}}\simeq$10~kpc, and a second one
at an [Fe/H]$\simeq$--0.3 outside the solar circle. More recent work based on
high resolution compilations of OC data \citep{sestito} find a steeper slope of
--0.17$\pm$0.01~dex~kpc$^{-1}$ within R$_{\rm{GC}}<$14~kpc, which still holds
when considering only the 10 clusters analyzed homogeneously by that group. A
sort of bimodal distribution is observed in their Figure~9, where between the
very steep slope of the inner clusters and the plateau of the outer ones there
is a small gap almost devoid of OC (9$<$R$_{\rm{GC}}$$<$12~kpc).

Our results are plotted in Figure~\ref{tr_rgc}, where we consider the trend of
[Fe/H] and [$\alpha$/Fe] with R$_{\rm{GC}}$. We do see a distinct slope in the
inner clusters and a flattening out for the outer ones. However, our sample
contains $\sim$15 more OC than the one of \citet{sestito}, and most of them
(including our five determinations) fall in the gap around
9$<$R$_{\rm{GC}}$$<$12~kpc, discussed above. With the addition of these clusters
we find a gentler slope of --0.05$\pm$0.01~dex~kpc$^{-1}$, in good agreement with
previous works \citep{fri95,friel02} and with the Disk Cepheids within
R$_{\rm{GC}}$$\simeq$11~kps \citep{and04,lem08}. If we exclude the clusters
outside R$_{\rm{GC}}$=12~kpc and remove the low resolution OC from
\citet{friel02}, the slope does not steepen significantly, becoming
--0.06$\pm$0.02~dex~kpc$^{-1}$. Also, the flattening out does not seem so abrupt
as in Figure~9 by \citet{sestito}. The paucity of OC in the flat part of the
relation (we have only Be~20, Be~22, Be~29 and Saurer~1 in our compilation)
surely calls for more high resolution studies, since as the data stand now, they
look compatible with both a plateau and a gradual change in slope. We would like
to note that NGC~2420 (already discussed in Section~\ref{lit-2420}) was placed
at [Fe/H]=--0.57~dex by \citet{smi87}, based on R$\simeq$16000 spectra, while we
find --0.05, in much better agreement with the global Galactic trend. This goes
in the direction of filling the gap in the \citet{sestito} compilation, and also
in the \citet{tw97} data-set, pointing more towards a gentle and continuous
decrease of [Fe/H].

The trend of $\alpha$-enhancement with R$_{\rm{GC}}$ is also of some importance,
since it unveils the role of SNe type Ia and II and their relative
contributions. \citet{yong05} found a tendency of the $\alpha$-enhancement to
increase with R$_{\rm{GC}}$, as did \citet{mag09}, who found this tendency in
good agreement with their chemical evolution model. Also model A by
\citet{chi01} predicted an increase of [$\alpha$/Fe] with R$_{\rm{GC}}$. In our
compilation, the trend appears as a weak slope, that is still perfectly
compatible with a flat distribution at the 1~$\sigma$ level. This, together with
the study of slope changes with time, is one typical case in which a
high quality, homogeneous analysis of $\sim$100 OC could give a clear and
definitive answer. 

%%%%%%%%%%%%%%%%%%%%%%%%%%%%%%%%%%%%%%%%%%%%%%%%% AMR \begin{figure} \centering
%%%%%%%%%%%%%%%%%%%%%%%%%%%%%%%%%%%%%%%%%%%%%%%%% AMR

\subsection{Trends with Age}

In spite of the fact that all models predict an evolution of Disk metallicity
with time, albeit maybe only in the first Gyrs, and the fact that such a
variation is observed in Disk stars \citep{red03,ben04}, there appears to be no
correlation at all between old OC abundances and ages \citep[see the review
by][]{fri95}. More recent results did not change this picture substantially. If
confirmed, the lack of an AMR in OC would point towards a different source of
chemical enrichment for OC stars \citep{yong05} with respect to the Disc stars.
In substance, the metallicity of OC stars seems to be more determined by the
location in which they formed, than by the time at which they formed.

What we find here is quite encouraging, although still not statistically
significant. We recall that our compilation includes 57 high resolution
abundance determinations, plus a handful of low resolution determinations by
\citet{friel02}. Although we have made no attempt to homogeneize the data,
except for a --0.16~dex correction to the low resolution abundances, this sample
is $\sim$50\% larger than any compilation presented before \citep[see
e.g.,][]{sestito,des09,mag09} and shows that the community is proceeding fast in
filling up the gaps of our knowledge of OC. Figure~\ref{tr_age} shows indeed a
weak trend of decreasing [Fe/H] and increasing [$\alpha$/Fe] with increasing
age. The slopes are very gentle at best and they are still compatible with no
trends at all. Nevertheless, if such trends exist, we can put some constraints
on them: for [Fe/H], the gradient should not be significantly larger than
--2.6$\pm$1.1~10$^{-11}$~dex~Gyr$^{-1}$ and for [$\alpha$/Fe] no larger than
1.1$\pm$5.0~10$^{-11}$~dex~Gyr$^{-1}$.

%__________________________________________________________________

\section{Summary and Conclusions}
\label{sec-sum}

We have analyzed high resolution spectra of three red clump giants in five OC,
three of them lacking any previous high resolution based chemical analysis.
Given the paucity of literature data, such a small sample is enough to increase
the whole body of high resolution data for OC by $\simeq$5\%. To compare our
results with the literature, we have compiled chemical abundances based on
high resolution data of 57 clusters from the literature. Given the recent and
fast progress in the field, this sample is $\sim$50\% larger than previous
literature compilations \citep[e.g.,][]{friel02,sestito,mag09}. The main results
drawn by the analysis of our five clusters are:

\begin{itemize} 
\item{We provide the first high resolution based abundance analysis of Cr~110
([Fe/H]=+0.03$\pm$0.02 ($\pm$0.10)~dex), NGC~2099 ([Fe/H]=+0.01$\pm$0.05
($\pm$0.10)~dex) and NGC~2420 ([Fe/H]=--0.05$\pm$0.03 ($\pm$0.10)~dex), which only
had low resolution determinations and the R$\simeq$16000 analysis by
\citet{smi87}; our new determination of the metallicity of NGC~2420 puts this
cluster in much better agreement with the global Galactic trends;} 
\item{The abundances found for NGC~7789 ([Fe/H]= +0.04$\pm$0.07
($\pm$0.10)~dex) and M~67 ([Fe/H]= +0.05$\pm$0.02 ($\pm$0.10)~dex) are in good
agreement with past high resolution studies;} 
\item{We provide the first high resolution based radial velocity
determination for Cr~110 ($<V_r>$=41.0$\pm$3.8~km~s$^{-1}$);} 
\item{We found that all our abundance ratios, with few exceptions generally
explained with technical details of the analysis procedure, are near-solar, as
is typical for OC with similar properties; we found solar ratios also for Na,
that is generally found overabundant, and for O, which is generally found
underabundant;} 
\item{We do not find any significant sign of anti-correlation (or correlation)
among Na, Al, Mg and O, in general agreement with past and recent results, and
we can say that if such correlations indeed are present in OC, they must be much
less extended than in Globular Clusters, amounting to no more than 0.2--0.3~dex at
most;}   
\end{itemize}

With our compilation of literature data, we also could examine global Galactic
trends, that are extremely useful to construct chemical evolution models for the
Galaxy in general and the Galactic Thin Disc in particular. For the metallicity
gradient we found a slope of  --0.06$\pm$0.02~dex~kpc$^{-1}$ considering only
the high resolution data within $R_{\rm{GC}}$=12~kpc. Our compilation contains
data, including our own determinations, that fill the small gap around
9$<$R$_{\rm{GC}}<$12 and point towards a gentle and continuous decrease, rather
than a two step drop such as in \citet{tw97} or a steep slope such as in
\citet{sestito}. We do find a flattening at $R_{\rm{GC}}>$12~kpc and a hint of
an increasing [$\alpha$/Fe] towards the outer Disk. Concerning the AMR, we do
not find any strong evidence for it, and we just note some very mild trends. If
an AMR is indeed present among OC, it must be very weak and we provide upper
limits to its slope both in [Fe/H] and [$\alpha$/Fe].

%______________________________________________________________

\begin{acknowledgements}

We warmly thank A. Bragaglia and A. Mucciarelli for their useful comments and
suggestions.We thank M. Tosi and D. Romano for their insights on the chemical
evolution modelling of the Milky Way and its Disks. We also warmly thank the
Calar Alto Support staff for their hospitality and a good time together. RC, CG
and EP acknowledge support by the Spanish Ministry of Science and Technology
(Plan Nacional de Investigaci\'on Cient\'{\i}fica, Desarrollo, e Investigaci\'on
Tecnol\'ogica, AYA2004-06343). EP acknowledges support from the Italian MIUR
(Ministero dell'Universit\'a e della Ricerca) under PRIN  2003029437,
"Continuities and Discontinuites in the Formation of the Galaxy." RC
acknowledges funding by the Spanish Ministry of Science and Innovation under the
MICINN/Fullbright post-doctoral fellowship program.

\end{acknowledgements}

%______________________________________________________________

\end{document}